\documentclass[11pt,reqno,oneside]{amsart}

\usepackage[margin=1in]{geometry}

\usepackage{amsmath,amssymb,amscd,amsthm}
\usepackage{mathrsfs}
\usepackage{mathtools}
\usepackage[dvips]{graphics}
\usepackage{enumitem}
\usepackage{mdframed}
\usepackage{cite}
\usepackage{comment}
\usepackage[colorlinks,pdfstartview=FitB]{hyperref}
\hypersetup{allcolors=blue}
\usepackage{graphicx}
\usepackage{xcolor}

\usepackage{todonotes}

\usepackage{tikz}
\usetikzlibrary{calc,
	matrix,
	fadings,
	fit,
	math,
	arrows.meta}
\usepackage{pgfplots}
\pgfplotsset{compat=1.18}

\usepackage{bbm}

\def\idty{\mathbbm{1}} 

\definecolor{purple}{rgb}{.5,0,1}
\definecolor{orange}{rgb}{1,.5,0}
\definecolor{pink}{rgb}{1,0,.5}

\numberwithin{equation}{section}
\newtheorem{theorem}{Theorem}[section]

\newtheorem{lemma}[theorem]{Lemma}
\newtheorem{corollary}[theorem]{Corollary}
\newtheorem{definition}[theorem]{Definition}
\newtheorem{remark}[theorem]{Remark}

\newtheorem*{bel*}{Belief}

\DeclareMathOperator{\Tr}{Tr}

\renewcommand\H{\mathcal{H}}

\newcommand\R{\mathbb R}
\newcommand\N{\mathbb N}
\newcommand\C{\mathbb C}
\newcommand\Z{\mathbb Z}

\newcommand\T{\mathbb{T}}

\newcommand\cF{\mathcal{F}}

\newcommand\vphi{\varphi}

\newcommand\beq{\begin{equation}}
\newcommand\eeq{\end{equation}}

\renewcommand{\Re}{{\mathrm{Re} \,}}
\renewcommand{\Im}{{\mathrm{Im} \,}}

\newcommand{\SU}{{\mathbb{SU}}}

\allowdisplaybreaks \numberwithin{equation}{section}

\makeatletter
\def\subsection{\@startsection{subsection}{2}%
	\z@{.5\linespacing\@plus.7\linespacing}{.5\linespacing}%
	{\normalfont\scshape\centering}}
\makeatother

\hypersetup{
	pdftitle = {Velocity bounds in periodic quantum walks},
	pdfauthor = {},
	pdfsubject = { },
	pdfkeywords = {}
}

\makeatletter
\renewcommand{\section}{\@startsection
  {section}{1}{\z@}%
  {-2.5ex \@plus -1ex \@minus -.2ex}%
  {1ex \@plus .2ex}%
  {\normalfont\Large\bfseries}}
  
\renewcommand{\l@section}{%
  \@tocline{1}{3pt plus 1pt}{0pt}{2.6em}{\normalsize\bfseries}%
}

\renewcommand{\subsection}{\@startsection
  {subsection}{2}{\z@}%
  {-2ex \@plus -1ex \@minus -.2ex}%
  {0.8ex \@plus .2ex}%
  {\normalfont\large\bfseries}}
  
  \renewcommand{\subsubsection}{\@startsection
  {subsubsection}{3}{\z@}%
  {-1.5ex \@plus -1ex \@minus -.2ex}%
  {0.6ex \@plus .2ex}%
  {\normalfont\bfseries}}
\makeatother

\newcommand{\LR}{\mathrm{LR}}
\newcommand{\Asy}{\mathrm{Asy}}

\begin{document}

\title{Lieb--Robinson Bounds for Electric Quantum Walks}

\author[H.\ Abdul-Rahman]{Houssam Abdul-Rahman}
\address{[H.\ Abdul-Rahman] Department of Mathematical Sciences, United Arab Emirates University, AL Ain, UAE}
\email{\href{mailto:houssam.a@uaeu.ac.ae}{houssam.a@uaeu.ac.ae}}

\author[C.\ Cedzich]{Christopher Cedzich}
\address{[C.\ Cedzich] Fakult\"at f\"ur Mathematik und Informatik, FernUniversit\"at in Hagen, Universit\"atsstr. 1, 58097 Hagen, Germany}
\email{\href{mailto:christopher.cedzich@fernuni-hagen.de}{christopher.cedzich@fernuni-hagen.de}}

\author[G. Stolz]{G\"unter Stolz}
\address{[G. Stolz] Department of Mathematics, 
University of Alabama at Birmingham, 
Birmingham, AL 35294 USA}
\email{\href{mailto: stolz@uab.edu}{stolz@uab.edu}}

\maketitle
\begin{abstract}
We derive Lieb--Robinson type bounds with optimal propagation velocity for shift-coin and split-step quantum walks in rational electric fields.  The velocity decays exponentially fast in the period of the field, showing long intervals of suppressed spreading. The proof rests on bounds for the imaginary part of the analytically continued dispersion relation of the electric walk. We show that the infimal Lieb--Robinson velocity coincides with the asymptotic velocity for these walks, while a nonhomogeneous example demonstrates that the two types of velocity need not agree in general. We also obtain an exponential tail estimate on the probability of a walk to lie outside the effective light cone.
\end{abstract}

\section{Introduction}

Quantum walks originally arose in the physics literature as a quantum-mechanical analogue of classical random walks \cite{QD-1,ambainisOnedimensionalQuantumWalks2001}. It was found that in the translation invariant setting quantum walks exhibit ballistic transport \cite{grimmettWeakLimitsQuantum2004,AhlbrechtEtal2011,DamanikSpreading2016} as opposed to the diffusive behavior of classical random walks and, in particular, Markov chains. This quadratically faster spreading provides a computational advantage in the design of quantum algorithms, for example, in search algorithms \cite{VenegasAndraca2012,Portugal2013}.

At the same time, quantum walks can be understood as the one-particle sector of particle number preserving quantum cellular automata (QCAs), see, e.g.,\ \cite{arrighiOverviewQuantumCellular2019,farrellyReviewQuantumCellular2020,schumacherReversibleQuantumCellular2004,vogtsDiscreteTimeQuantum2009}. As such, they furnish a lattice model of quantum dynamics in which time proceeds iteratively in discrete steps. This point of view led to the study of characteristics of quantum walks such as their transport phenomena or their spectral properties, e.g., in random \cite{AhlbrechtVolkherScholzWerner2011,HamzaJoyeStolz2009,Joye2012} or quasi-periodic setups \cite{FOZ2016,CFO1,CFLOZ}.

In the current manuscript we are specifically interested in finding optimal Lieb--Robinson transport bounds (in the sense of minimal velocity) for one-dimensional quantum walks in external homogeneous rational electric fields. These models were introduced in \cite{Cedzichetal2013} where it was found that the dynamics depends sensitively on the arithmetic properties of the field: rational fields entail ballistic motion with revivals on time scales of the order of the denominator of the field, that is, almost perfect restorations of the initial state. For well-approximable irrational fields, better and better revivals alternate with farther and farther excursions leading to a dynamical signature of singular continuous spectrum; see also \cite{CedzichPhD} for a Gordon-type argument for Liouville fields. Typical (irrational) fields lead to Anderson localization, that is, pure point spectrum with exponentially decaying eigenfunctions \cite{CedzichWerner2021}.

Our work can be seen as a continuation of a more long term project on identifying periodic quantum systems with absolutely continuous spectrum and {\it slow} ballistic transport. Velocity bounds have been obtained in  \cite{ADFS24} and, more generally, in \cite{ARFFL25} for discrete Schr\"odinger operators with periodic potentials at large coupling. Related general bounds on the maximal group velocity of quantum walks with periodic coins were obtained in \cite{ARJO26}.

On the other hand, the present manuscript is also a continuation and, in some sense, completion of the works \cite{ARS2023-CMP,ARCSW25} where the {\it asymptotic velocity} of quantum walks in rational electric fields were considered. As our main result here shows, in this setting, the asymptotic velocity and the Lieb--Robinson velocity are identical. This naturally leads to the question whether this equality holds in more general scenarios. We show that this cannot be true in full generality via a non-homogeneous counterexample which localizes and thus has vanishing asymptotic velocity but whose Lieb--Robinson velocity is strictly positive. This leaves the class of homogeneous walks as a natural candidate where we believe that this equality holds.
We do not offer a proof, but we derive an exponential bound on the probability for being outside of the classical light cone defined by the asymptotic velocity for the model considered here. This bound is known for general translation invariant quantum systems \cite{CJWW25}, however, our argument works with the Lieb--Robinson velocity instead of the asymptotic velocity.

As a final reason for our interest in Lieb--Robinson velocities, we stress their importance in quantum many-body theory and, specifically, in the study of quantum lattice systems, where they are currently the sole well-defined concept that allows to measure group velocities, see, e.g., 
\cite{LiebRob,HasKoma,NachSims}. 
Thus, we think that even beyond the actual result the mathematical techniques for deriving one-particle analogs of Lieb--Robinson bounds are important.
\smallskip

We now outline the contents of the body of this paper. Section~\ref{sec:prelims} has review character: in Section~\ref{sec:basics} we recall the basics of one-dimensional quantum walks $W$ with an internal degree of freedom and in Section~\ref{sec:LRvel} we introduce the concepts of Lieb--Robinson velocities. Importantly, we introduce the infimal Lieb--Robinson velocity $v_{\LR}^{\inf}(W)$ and prove Lemma~\ref{scaling} which allows to scale Lieb--Robinson velocities with time. We also recall the definition of the asymptotic (group) velocity $v_{\Asy}(W)$ and find that in general $v_{\Asy}(W) \leq v_{\LR}^{\inf}(W)$ for quantum walks, see Lemma~\ref{lem:vAsyvLR}. In Section \ref{sec:counterexample}, we exhibit an example for which this inequality is strict. Then, in Section~\ref{sec:Fourier}, we show how to Fourier-transform any translation invariant walk $W$ to its Floquet-Bloch matrix $\widehat{W}(\theta)$ and discuss its basic properties. 

In Section~\ref{sec:U} we introduce our first model, the shift-coin walk $U_\Phi$ in homogeneous electric field with rational phase $\Phi = q/p$, $q$ and $p$ relatively prime, see \eqref{def:U} to \eqref{def:Phi}. We discuss the Floquet-Bloch matrix of the translation invariant power $U_\Phi^p$, including Lemma~\ref{lem:trace} which is Lemma 2 in the Appendix of \cite{Cedzichetal2013}. This result is central to our work, as it is for the results in \cite{ARCSW25}, and one easily concludes the dispersion relation \eqref{disrel} of $U_\Phi^p$. This immediately allows to state Corollary~\ref{coro:measure} on the Lebesgue measure of the spectrum of $U_\Phi$, which decays exponentially in $p$. 

We then formulate our first main result, Theorem~\ref{thm:LR-bound}.  It is a Lieb--Robinson bound for $U_\Phi$ with rather explicit and, as it turns out, optimal control of the velocity and constants. As a consequence, we find that the infimal Lieb--Robinson velocity $v_{\LR}(U_\Phi)=|a|^p$ for $p$ odd and $v_{\LR}(U_\Phi)=|a|^{p/2}$ for $p$ even, where $a$ is the complex \emph{transmission parameter} in the coin matrix \eqref{def:C}. This coincides with the asymptotic velocity (Corollary~\ref{cor:vLReqvAsy}).
Theorem~\ref{thm:LR-bound} is proved in Sections~\ref{sec:Proof-odd} ($p$ odd) and \ref{sec:Proof-even} ($p$ even). The proof adapts the methods of \cite{ADFS24} and \cite{ARFFL25}, but achieving optimality requires additional technical refinements that make the argument more involved. 

As another consequence, we determine the tail-probability that the walk starting from position $j$ will stay in the effective light cone $|j-k|> v_{\varepsilon}|n|$, where $v_{\LR}^{\inf} + \varepsilon$, see Corollary~\ref{coro:tail}. In general, this probability is known to behave like $\exp[-c \varepsilon^{3/2}|n|]$, see \cite{CJWW25}, and we determine the constant $c$ in our explicit setting.

In Section~\ref{sec:W} we will outline how our main results extend to electric {\it split-step walks}, see the definition in \eqref{def:W} to \eqref{def:W-Phi}. This is Corollary~\ref{coro:W-LR}. Using the {\it sieving} technique introduced in Section~3.2 of \cite{ARCSW25} this can be reduced to shift-coin walks via Theorem~\ref{thm:LR-bound}.

\smallskip

While doing this work we understood that the general results in recent papers by Cedzich, Joye, Werner and Werner \cite{CJWW25} and by Sigal and Wu \cite{SW25} can also be used to derive  Lieb--Robinson bounds similar to the ones found here. Our methods differ from those used in these works and offer a complementary approach to establishing such bounds.

\section*{Acknowledgements} 

C. C. and H. A.-R. gratefully acknowledge the hospitality of the Mathematisches Forschungsinstitut Oberwolfach (MFO), where part of this work was carried out.\\
 G.\ S.\ is supported by Yulia and Newton. 

\section{Preliminaries} \label{sec:prelims}
\subsection{Basics} \label{sec:basics}
We consider the Hilbert space $\H = \ell^2(\Z;\C^2) = \ell^2(\Z) \otimes \C^2$ with orthonormal basis $|n,\sigma \rangle$, $n\in \Z$, $\sigma \in \{\pm\}$. We will frequently switch between viewing $\H$ as  $\ell^2(\Z;\C^2)$ and $\ell^2(\Z) \otimes \C^2$, respectively..

A linear operator $A$ on the finitely supported sequences $\ell_0(\Z;\C^2)$ is called \emph{banded} if there exists $R\in \N_0$ such that \begin{equation} \label{banded} 
A(j,k) = 0 \;\: \mbox{for} \:\: |j-k|>R.
\end{equation}
Here the matrix-elements of $A$ are defined as two-by-two blocks
\begin{equation} 
A(j,k) =\begin{bmatrix} \langle j,+|A|k,+ \rangle & \langle j,+|A|k,-\rangle \\ \langle j,-|A|k,+ \rangle & \langle j,-|A|k,- \rangle \end{bmatrix}.
\end{equation}
The \emph{band radius} of $A$ is the smallest $R$ such that \eqref{banded} holds and its \emph{band width} is $2R+1$.

It can be shown that a banded $A$ is closable with $\overline{A} = A_{\max}$, where 
\begin{equation} 
D(A_{\max}) = \{ \varphi \in \H: A\varphi \:\:\mbox{given by}\:\: (A\varphi)(j) = \sum_k A(j,k)\varphi(k) \:\: \mbox{is in} \:\:\H\} 
\end{equation}
and $A_{\max}\varphi = A\varphi$ for $\varphi\in D(A_{\max})$. We will identify $A= \bar{A}$.

A banded $A$ is bounded, $A\in B(\H)$, if and only if $\sup_{j,k} \|A(j,k)\| = C < \infty$. In this case $\|A\| \leq (2R+1)C$.

If $A$ is banded (and bounded), then $A^k$ is banded (and bounded). An upper bound for its band radius is given by $kR$ and for its band width by $2kR+1$.

\vspace{.3cm}

$W$ is said to be a \emph{quantum walk} in $\H$, if $W$ is in $U(\H)$ and banded.

\vspace{.3cm}

A banded $A$ is \emph{translation invariant} if $AT=TA$ for $T=T_1 \otimes \idty_{\C^2}$, where $(T_1 \psi)(k) = \psi(k-1)$. This is equivalent to 
\begin{equation} 
A(j+n,k+n) = A(j,k) \:\: \mbox{for all $j, k, n \in \Z$}. 
\end{equation}
A translation invariant banded $A$ is bounded and  can be represented as
\begin{equation} \label{Wrep}
A = \sum_{\ell}A(\ell,0) T^{\ell},
\end{equation}
where the sum is finite due to bandedness.
\subsection{Lieb--Robinson velocities} \label{sec:LRvel}
\begin{definition}
Let $W$ be a quantum walk on $\ell^2(\Z;\C^2)$. We say that $v\geq 0$ is a \emph{Lieb--Robinson velocity} (LR-velocity) for $W$ if there exist constants $C<\infty$ and $c>0$ such that 
\begin{equation} \label{LRvel}
\|W^n(j,k)\| \leq C e^{-c(|j-k|-v|n|)}. 
\end{equation}
for all $j, k \in \Z$ and $n\in \Z$.
\end{definition}

The bound \eqref{LRvel} implies the more familiar commutator version of the Lieb--Robinson bound from many-body theory, e.g.\ \cite{HasKoma} or \cite{NachSims}. The following is adapted from \cite{ADFS24}. Let $A, B \in B(\ell^2(\Z;\C^2))$ and consider the {\it local} operators $A_j = P_j A P_j$, $B_k = P_k B P_k$, where $P_j$ is the orthogonal projection onto span$\{|j,+\rangle, |j,-\rangle \}$. Suppose that \eqref{LRvel} holds. Let $\tau_n(C) = W^{-n} C W_n$ be the discrete time Heisenberg evolution of $C\in B(\ell^2(\Z;\C^2))$. Then
\begin{eqnarray} \label{commbound}
\| [ \tau_n(A_j), B_k ] \| & \le & \|A_j W^n B_k\| + \|A_j^* W_n B_k^*\| \notag  \\
& \le & 2\|A_j\| \|B_k\| \|P_j W^n P_k \|.
\end{eqnarray}
It is elementary to see that 
\begin{equation}  \label{Normfact} 
\|P_j W^n P_k\| = \|W^n(j,k)\|.
\end{equation}
Thus \eqref{commbound} and \eqref{LRvel} yield the bound
\begin{equation} \label{commbound2}
\| [ \tau_n(A_j), B_k ] \| \le  2C \|A_j\| \|B_k\| e^{-c(|j-k|-v|n|)}.
\end{equation}

Moreover, if $W$ has band radius $R$, then $R$ is trivially a LR-velocity for $W$. This follows from two facts: 

(i) $\|W^n(j,k)\|\leq 1$ for all sites $j$ and $k$, because of \eqref{Normfact}.

(ii) $R|n|$ is the band radius of $W^n$ and thus $\|W^n(j,k)\|=0$ for $|j-k| > R|n|$. 

Combined this implies $\|W^n(j,k)\| \leq e^{- (|j-k|-R|n|)}$.

\begin{lemma} \label{scaling}
Let $W$ be a quantum walk and $p$ a positive integer. Then $v\geq 0$ is a LR-velocity for $W$ if and only if $pv$ is a LR-velocity for $W^p$.
\end{lemma}

\begin{proof} The ``only if'' part is trivial. For the ``if'' part we will show that if $v$ is a LR-velocity for $W^p$, then $v/p$ is a LR-velocity for $W$.

The assumption means that there exist $C<\infty$ and $c>0$ such that
\begin{equation} 
\|W^{np}(j,k)\| \leq C e^{-c(|j-k|-nv)} \:\: \mbox{for all}\:\: j, k \in \Z, \: n \in \N. 
\end{equation}
Let $m\in \N$ and write $m=np+s$ with $n\in \Z$ and $s \in \{0,\ldots,p-1\}$. Then $W^m = W^{np} W^s$. Let $R$ be the band radius of $W$. Then
\begin{equation} \label{Step1}
\|W^m(j,k)\| \leq \sum_{\ell=k-Rs}^{k+Rs} \|W^{np}(j,\ell)\| \|W^s(\ell,k)\|.
\end{equation}
Since $W$ is unitary, so is $W^s$, and hence $\|W^s(\ell,k)\|\leq 1$. Moreover,
\begin{equation}
|j-\ell| \geq |j-k|-|k-\ell| \geq |j-k|-R(p-1).
\end{equation}
Using these observations together with the assumption, we obtain
\begin{align}
\|W^m(j,k)\| &\leq \sum_{\ell=k-Rs}^{k+Rs}   C e^{-c\left(|j-k|-vn-R(p-1)\right)}    \nonumber\\
&=(2Rs+1)C    e^{-c\left(|j-k|-\frac{v}{p}m\right)}   e^{-cv\frac{s}{p}}   e^{cR(p-1)}    \nonumber\\
&\leq (2R(p-1)+1)C e^{cR(p-1)}   e^{-c\left(|j-k|-\frac{v}{p}m\right)}. \label{Step2}
\end{align}

\end{proof}

We define the \emph{infimal} LR-velocity of $W$ as
\begin{equation} \label{infvLR}
v_{\LR}^{\inf}(W) := \inf \{v\geq 0: v \:\:\mbox{is a LR-velocity for $W$}\}.
\end{equation}

Obviously, if $v$ is a LR-velocity for $W$, any $v'>v$ is also a LR-velocity for $W$. Thus the set of all LR-velocities of $W$ is either the open half-line $(v_{\LR}^{\inf}(W), \infty)$ or the closed half-line $[v_{\LR}^{\inf}(W), \infty)$.

In particular, the infimum need not, a priori, be a LR-velocity. Appendix~\ref{sec:vAsyvLR} gives an example in which the infimum is attained; whether the open alternative can occur remains unclear.

We next compare $v_{\LR}^{\inf}(W)$ with the asymptotic propagation velocity. Let $W$ be a banded quantum walk on $\ell^2(\Z;\C^2)$, and let $Q$ denote the position operator
\begin{equation}
(Q\psi)(n)=n\psi(n),
\qquad
D(Q)
=
\left\{
\psi\in\ell^2(\Z;\C^2):
\sum_{n\in\Z}n^2|\psi(n)|^2<\infty
\right\}.
\end{equation}
Its discrete-time Heisenberg evolution with respect to some walk $W$ is
\begin{equation}
Q(n)=W^{-n}QW^n,
\qquad n\in\Z.
\end{equation}
Since $W$ is banded, it preserves $D(Q)$ and hence $D(Q(n))=D(Q)$. For every normalized $\psi\in D(Q)$, define the velocity of $\psi$ with respect to $W$ as
\begin{equation}\label{vAsy1}
v(W,\psi)
=
\limsup_{n\to\infty}
\frac{1}{n}\|Q(n)\psi\|,
\end{equation}
and the \emph{asymptotic velocity} as
\begin{equation}\label{vAsy2}
v_{\Asy}(W)
=
\sup_{\substack{\psi\in D(Q)\\ \|\psi\|=1}}
v(W,\psi).
\end{equation}

Every LR-velocity for $W$ bounds the asymptotic velocity from above. As shown in Appendix~\ref{sec:vAsyvLR}, this follows by adapting the continuous-time argument of \cite[Theorem~A.1]{ADFS24} to discrete-time quantum walks. In particular, 
\begin{equation}\label{vAsyvLR}
v_{\Asy}(W)\leq v_{\LR}^{\inf}(W),
\end{equation}
see Lemma \ref{lem:vAsyvLR}.

\subsection{Fourier transform} \label{sec:Fourier}
Let $\cF: \ell^2(\Z; \C^2) \to L^2([0,2\pi);\C^2) = L^2(\T;\C^2)$, with the torus $\T = \R/2\pi\Z$, be the component-wise Fourier transform, i.e.\ $\cF = \cF_0 \otimes \idty_{\C^2}$ with
\begin{equation} \label{Fourier} (\cF_0 \psi)(\theta) = \frac{1}{\sqrt{2\pi}} \sum_k e^{ik\theta} \psi(k), \:\: \psi \in \ell^2(\Z), 
\end{equation}
\begin{equation} \label{Fourierinf}
(\cF_0^{-1} \vphi)(n) = \frac{1}{\sqrt{2\pi}} \int_0^{2\pi} e^{-in\theta} \vphi(\theta)\,d\theta, \:\: \vphi \in L^2(\mathbb{T}). 
\end{equation}
If $W$ is translation invariant, then 
\begin{equation}  \label{hatWdef}
\cF W \cF^{-1} = \widehat{W},
\end{equation} where $\widehat{W}$ is the generalized multiplication operator by $\widehat{W}(\theta) \in U(2)$, $\theta \in \T$,
\begin{equation} 
(\widehat{W} \psi)(\theta) = \widehat{W}(\theta) \psi(\theta) \:\: \mbox{for all $\psi \in L^2(\T; \C^2)$}.
\end{equation}
Here the unitary $2\times 2$ matrix $\widehat{W}(\theta)$ is given by
\begin{equation} \label{hatW}
 \widehat{W}(\theta) = \sum_{\ell} W(\ell,0) e^{i\ell \theta}. 
 \end{equation}
This is a consequence of \eqref{Wrep} and $\cF_0 T \cF_0^{-1} = M_e$, the maximal multiplication operator with $e(\theta)=e^{i\theta}$. We call $\widehat{W}(\theta)$ the \emph{Floquet matrix} of $W$ and note that its entries are trigonometric polynomials since $W$ is banded.

By standard arguments the spectrum of $W$ can be expressed as the union of the spectra of the fibers $\widehat{W}(\theta)$,
\begin{equation} \label{spectrumW}
\sigma(W) = \bigcup_{\theta \in \T} \sigma(\widehat{W}(\theta)).
\end{equation}

For a translation invariant walk it holds that
\begin{equation} \label{Bloch}
W(j,k) = \frac{1}{2\pi} \int_0^{2\pi} \widehat{W}(\theta) e^{-i(j-k)\theta}\,d\theta
\end{equation}
(calculate $\cF^{-1} \widehat{W} \cF$), which can be considered as the inverse of \eqref{hatW} or the Bloch transform of $W$. Recall that \eqref{Bloch} is $2\times 2$-matrix-valued.

If $W_1$ and $W_2$ are translation invariant quantum walks, then \eqref{hatWdef} implies that
\begin{equation} \label{product}
\widehat{W_1W_2}(\theta) = \widehat{W_1}(\theta) \widehat{W_2}(\theta). 
\end{equation}
This gives 
\begin{equation} \label{nPower}
\widehat{W^n}(\theta) \psi(\theta) = (\widehat{W})^n(\theta) \psi(\theta)
\end{equation}
for $n\in \N$. A more general functional calculus could be built here. For example, elementary considerations show that if $W$ is a translation invariant quantum walk, then so is $W^{-1}$ and the Floquet matrices are related by 
\begin{equation} \label{Floquetinv}
\widehat{W^{-1}}(\theta) = (\widehat{W}(\theta))^{-1},
\end{equation}
so that \eqref{nPower} holds for all $n\in \Z$.

Thus the Fourier transform partially diagonalizes $W$. One can further diagonalize $\widehat{W}$ pointwise by
\begin{equation} \label{hatWdiag} 
\widehat{W}(\theta) = \sum_{\sigma\in {\pm}} e^{i\omega_{\sigma}(\theta)} P_{\sigma}(\theta),
\end{equation}
where the $\omega_{\sigma}(\theta)$ are called \emph{dispersion relations} of $\widehat{W}(\theta)$ and the $P_{\sigma}(\theta)$ are corresponding rank-one orthogonal eigenprojections. Both $\omega_{\sigma}$ and $P_{\sigma}$ can be chosen real-analytic in $\theta \in \T$ \cite{AhlbrechtEtal2011,ahlbrechtAsymptoticBehaviorDecoherent2013}.

If $W$ is translation invariant, it is known that 
\begin{equation}
\lim_{n\to\infty} \frac{1}{n} Q(n) \psi = V\psi, \quad \text{ where }V=
\mathcal{F}^{-1}\left[\int_{\mathbb{T}}^\oplus\sum_{\sigma\in\pm}\omega'_\sigma(\theta)P_\sigma(\theta)d\theta\right]\mathcal{F}
\end{equation}
for $\psi \in D(Q)$, see \cite{ahlbrechtAsymptoticEvolutionQuantum2011} and also \cite{DamanikSpreading2016,CJWW25}.
Since the $D(Q)$ is dense in $\ell^2(\Z;\C^2)$, the dispersion relations determine the asymptotic velocity in the translation invariant setting via
\begin{equation}\label{vAsy-Formula}
v_{\Asy}(W)=\|V\|_{\ell^2(\Z;\C^2)}=\max_{\substack{\sigma\in\{+,-\}\\ \theta\in\T}}\left|\omega'_\sigma(\theta)\right|.
\end{equation}

\section{The models and main results}\label{sec:Models}

This section introduces two closely related one-dimensional unitary models in a rational electric field. We begin with the electric shift--coin walk (see Section \ref{sec:U}), which serves as the basic model, and review its Fourier--Floquet representation and the known formula for its asymptotic velocity. Building on this analysis, we derive an exact formula for the Lebesgue measure of the spectrum and an explicit one-parameter family of LR-velocities, together with a quantitative tail bound outside the resulting light cone. We then transfer the Lieb--Robinson estimate to electric split-step walks through the sieving relation (see Section \ref{sec:W}). 

\subsection{Electric shift-coin walks} \label{sec:U}
The \emph{shift-coin walk} is the unitary
\begin{equation} \label{def:U}
U = SC, 
\end{equation}
where the \emph{conditional shift} $S$ is given by
\begin{equation} \label{def:S}
S|n,+\rangle = |n+1,+\rangle, \qquad S|n,-\rangle = |n-1,-\rangle,
\end{equation} 
for all $n\in\Z$, and the \emph{coin} $C$ is a unitary that acts locally at each site. We are especially interested in the translation invariant case in which $C$ acts identically at every site, that is, $(C\vphi)(n) = C(0) \vphi(n)$ for all $\vphi \in \ell^2(\Z;\C^2)$ and $n\in\Z$ with $C(0)\in\SU(2)$ given by
\begin{equation} \label{def:C}
C(0) =\begin{bmatrix} a & b \\ - \bar{b} & \bar{a} \end{bmatrix}, \qquad |a|^2 + |b|^2 = 1. 
\end{equation} 
Then the shift-coin walk $U=SC$ is translation invariant with Fourier representation
\begin{equation}\label{eq:U0Fourier}
\widehat{U}(\theta) =\begin{bmatrix} ae^{i\theta} & b e^{i\theta} \\ -\bar{b} e^{-i\theta} & \bar{a} e^{-i\theta} \end{bmatrix}, 
\end{equation}
as follows from \eqref{hatW} and \eqref{product}.

The shift-coin walk in a homogeneous static electric field $\Phi \in [0,1)$ is defined by
\begin{equation} \label{def:U-Phi} 
U_{\Phi} = F_{\Phi} U. 
\end{equation}
Here $F_{\Phi} = e^{2\pi i\Phi Q}$, where $(Q\psi)(n) = n \psi(n)$, $n\in \Z$, is the \emph{position operator} in $\ell^2(\Z;\C^2)$. Both $U$ in \eqref{def:U} and $U_{\Phi}$ in \eqref{def:U-Phi} have band radius 1. Unless $\Phi=0$, $U_{\Phi}$ is not translation invariant. 

We focus on the case of a \emph{rational field}, which means that we set
\begin{equation} \label{def:Phi}
\Phi = \frac{q}{p}, \:\: q, p \in \N \:\:\mbox{relatively prime}, 
\end{equation}
so that $e^{2\pi i \Phi}$ is a primitive $p$-th root of unity.
Then $U_{\Phi}$ is $p$-periodic in the sense that its matrix elements satisfy
\begin{equation} 
U_{\Phi}(n+p,m+p) = U_{\Phi}(n,m) \:\: \mbox{for all $n,m\in \Z$}. 
\end{equation}
A direct calculation further shows that $U_\Phi^p$ is translation invariant, thereby allowing the use of Fourier methods in the analysis below.

We first review the Fourier transform of $U_\Phi^p$ and the resulting exact formula for the asymptotic velocity of $U_\Phi$, as established in \cite{ARCSW25}. We include the details needed for the new results below. First, the same Floquet analysis yields the Lebesgue measure of $\sigma(U_\Phi)$ and its exponential decay rate as $p\to\infty$; see Corollary \ref{coro:measure}. We then present our main result, a family of LR-velocities for $U_\Phi$, in Theorem \ref{thm:LR-bound}. Our presentation of the Fourier and Floquet analysis follows the approaches developed in \cite{CedzichPhD,Cedzichetal2013}.

Since $F_\Phi$ is a multiplication operator, its Fourier transform amounts to a shift $\theta\mapsto\theta+\Phi$ in the Fourier variable.
It follows that $(\widehat{U_{\Phi}}f)(\theta) = \widehat{SC}(\theta+2\pi \Phi) f(\theta+2\pi \Phi)$. 
Iterating this identity for $\Phi=q/p$ yields
\begin{equation} 
(\widehat{U_{\Phi}}^p f)(\theta) = \widehat{U^p_{\Phi}}(\theta) f(\theta), 
\end{equation}
with the unitary multiplication operator 
\begin{equation} \label{hatUPhip}
\widehat{U_\Phi^p}(\theta) := \begin{bmatrix} e^{2\pi i \Phi} & 0 \\ 0 & e^{-2\pi i \Phi} \end{bmatrix}\ \begin{bmatrix} a e^{i\theta} & b e^{i\theta} \\ -\bar{b} e^{-i\theta} & \bar{a} e^{-i\theta} \end{bmatrix}\cdots\begin{bmatrix} e^{2\pi i p\Phi} & 0 \\ 0 & e^{-2\pi i p\Phi} \end{bmatrix}\ \begin{bmatrix} a e^{i\theta} & b e^{i\theta} \\ -\bar{b} e^{-i\theta} & \bar{a} e^{-i\theta} \end{bmatrix}.
\end{equation}
From here on, we identify the torus with the complex unit circle and write $\widehat{U^p_{\Phi}}(e^{i\theta})$ for $\widehat{U^p_{\Phi}}(\theta)$. 
We note in passing that the Floquet matrix in \eqref{hatUPhip} is a trigonometric polynomial and can hence be analytically continued to a matrix-valued function of $z\in \C\setminus \{0\}$ which we accordingly denote by $\widehat{U_\Phi^p}(z)$.

For later purpose, we collect (some of) its properties: $\det \widehat{U_\Phi^p}(z)=1$ for all $z\in \C \setminus \{0\}$, therefore
\begin{enumerate}
\item $\widehat{U^p_{\Phi}}(z)$ is unitary iff $|z|=1$,
\item if $\widehat{U_\Phi^p}(z)$ is diagonalizable we have
\begin{equation}
\widehat{U_\Phi^p}(z)=\lambda_+(z)P_+(z) +\lambda_-(z)P_-(z)
\end{equation}
with eigenprojections $P_\pm(z)$ which are orthogonal if $|z|=1$,
\item its eigenvalues $\lambda_+(z)$ and $\lambda_-(z)$ are solutions of the secular equation
\begin{equation}
\lambda^2-\Tr \widehat{U_\Phi^p}(z)\lambda+1=0
\end{equation}
and hence are fully determined by $\Tr\left(\widehat{U_\Phi^p}(z)\right)$,
\item its eigenvalues reciprocals of each other, that is, $\lambda_+(z)=1/\lambda_-(z)$. We shall oftentimes write $\lambda_+(z)=e^{i\omega(z)}$ for some function $\omega:\C\to\overline{\C_+}$.
\end{enumerate}

Since $\widehat{U_\Phi^p}(e^{i\theta})$ is unitary and has determinant one, its eigenvalues, counted with multiplicity, are of the form $\lambda_\pm(e^{i\theta},a)=e^{\pm i\omega(e^{i\theta},a)}$ for a suitable \emph{dispersion relation} $\omega(e^{i\theta},a)\in[0,\pi]$ which are completely determined by its trace through
\begin{equation}\label{disrel}
\frac12\Tr\bigl(\widehat{U_\Phi^p}(e^{i\theta})\bigr)=\cos\bigl(\omega(e^{i\theta},a)\bigr).
\end{equation}

It turns out that the structure of the product in \eqref{hatUPhip} permits to give an analytic expression for the dispersion relations through the following trace identity from \cite[Lemma 2]{Cedzichetal2013}:
 \begin{lemma}\label{lem:trace}
		Let $p\in\mathbb{N}$, and let $\alpha$ be a primitive $p$-th root of unity. Consider the matrices
		\begin{equation}\label{CandR}
			A=\begin{bmatrix}a&b\\c&d\end{bmatrix} \quad\text{and}\quad
			R=\begin{bmatrix}\alpha&0\\0&\alpha^{-1}\end{bmatrix}.
		\end{equation}
		Then\footnote{A sign typo in \cite{Cedzichetal2013} is corrected here.}
		\begin{equation} \label{ExplOmega}
			\Tr\left( A RA R^2\cdots A R^p\right)=	\begin{cases} 	a^p+d^p, & $p$ \text{ odd},\\
										-\bigl(a^{p}+d^{p}\bigr)+2(-1)^{p/2}\bigl((ad)^{p/2}-\det(A)^{p/2}\bigr),& $p$ \text{ even}.
						\end{cases}
		\end{equation}
\end{lemma}
Writing $a=e^{i\phi}|a|$ and applying Lemma~\ref{lem:trace} to $A=\widehat{U}(e^{i\theta})$ in \eqref{eq:U0Fourier} and $\alpha=e^{2\pi i\Phi}$ gives
\begin{equation}\label{traceU}
\cos\bigl(\omega(e^{i\theta},a)\bigr)=\begin{cases}
|a|^p\cos\bigl(p(\theta+\phi)\bigr), & \text{$p$ odd},\\
\left(|a|^p\left((-1)^{p/2}-\cos\bigl(p(\theta+\phi)\bigr)\right)-(-1)^{p/2}\right),& \text{$p$ even}.
\end{cases}
\end{equation}
Plugging this into \eqref{vAsy-Formula} determines the asymptotic velocity of $U_\Phi$ explicitly. 
To state the result uniformly for odd and even periods, define
\begin{equation}\label{def:kappa}
\kappa(p)=
\begin{cases}
1, & p\text{ odd},\\
2, & p\text{ even}.
\end{cases}
\end{equation}
Differentiating the dispersion relations and maximizing over $\theta\in\T$ then gives  \cite[Lemma 3.1]{ARCSW25}
\begin{equation}\label{eq:vAsy-Up}
v_{\Asy}(U_\Phi^p)=\max_{\theta\in\T}|\partial_\theta\omega(e^{i\theta},a)|=p|a|^{p/\kappa(p)}.
\end{equation}

From the explicit expression for the dispersion relations in \eqref{traceU} one can also directly infer the Lebesgue measure of the spectrum.
The function $x\mapsto\cos^{-1}x$ is decreasing on $[-1,1]$. Moreover, for $0\leq x\leq1$,
$\cos^{-1}(2x-1)=2\cos^{-1}\sqrt{x}$ and $\cos^{-1}(1-2x)=2\sin^{-1}\sqrt{x}$. Combining these observations with \eqref{traceU}, we obtain the following ranges for $\pm\omega(e^{i\theta},a)$:
\begin{align}\label{eq:spec}
p\text{ odd}:\quad& \left[\cos^{-1}(|a|^p),\cos^{-1}(-|a|^p)\right]\cup \left[-\cos^{-1}(-|a|^p),-\cos^{-1}(|a|^p)\right] \notag\\
p/2\text{ even}:\quad& \left[2\cos^{-1}(|a|^{p/2}),\pi\right]\cup\left[-\pi,-2\cos^{-1}(|a|^{p/2})\right] \notag \\
p/2\text{ odd}:\quad&\left[-2\sin^{-1}(|a|^{p/2}),2\sin^{-1}(|a|^{p/2})\right].
\end{align}

For $p/2$ even, the two intervals meet at $\pi\bmod2\pi$ and therefore describe a single connected arc on $\T$. This is because for $\theta$ such that $\theta=-\phi+2\pi k/p$ for some $k\in\Z$ the walk $\widehat{U_\Phi^p}(e^{i\theta})$ has a doubly degenerate eigenvalue $-1$.
Similarly, for $p/2$ odd, the two branches of the dispersion relation meet at $\omega(e^{i\theta},a)=0$ whenever $\theta=-\phi+2\pi k/p$ for some $k\in\Z$ corresponding to the doubly degenerate eigenvalue $+1$ of $\widehat{U_\Phi^p}(e^{i\theta})$.

Combining these ranges with the invariance of $\sigma(U_\Phi)$ under rotation by $e^{2\pi i\Phi}$, see \cite{ARCSW25}, shows that, when $0<|a|<1$, the spectral arcs have the following structure:
\begin{itemize}
\item[(i)] If $p$ is odd, then $\sigma(U_\Phi)$ consists of $2p$ disjoint arcs, each of length $\frac{2}{p}\sin^{-1}(|a|^p)$.
\item[(ii)] If $p$ is even, then $\sigma(U_\Phi^p)$ is a single connected arc. In the even sub-case, the length of this arc is
$4\sin^{-1}(|a|^{p/2})$.
Hence $\sigma(U_\Phi)$ consists of $p$ disjoint arcs, each of length
$\frac{4}{p}\sin^{-1}(|a|^{p/2})$.

\end{itemize}

We have therefore proved the following result.

\begin{corollary}\label{coro:measure}
Let $U_\Phi$ be the electric shift--coin walk with rational field $\Phi=q/q$ defined in \eqref{def:U-Phi}, with $a\in\C$ and $0\leq|a|\leq1$. Then the Lebesgue measure (arc length) of the spectrum of $U_\Phi$ is
\begin{equation}\label{eq:spectral-measure}
|\sigma(U_\Phi)|=4\sin^{-1}(|a|^{\frac{p}{\kappa(p)}}),
\end{equation}
where $\kappa(p)$ is defined in \eqref{def:kappa}.
\end{corollary}

Note that for fixed $0<|a|<1$, Corollary~\ref{coro:measure} in conjunction with $\sin^{-1} x=x+\mathcal{O}(x^3)$  implies that the Lebesgue measure of the spectrum of $U_\Phi$ decays exponentially in the period $p$ of the field. 

We now turn to the corresponding estimate for the LR-velocity by providing an explicit one-parameter family of LR-velocities for $U_\Phi$: 

\begin{theorem}\label{thm:LR-bound}
Let $0<|a|<1$ and $U_\Phi$ be the electric shift-coin quantum walk with rational field $\Phi=q/p$ defined in \eqref{def:U-Phi}. 
Then, with $\kappa(p)$ as in \eqref{def:kappa}, for every $\sigma$ satisfying
\begin{equation}\label{eq:sigma-range}
0<\sigma<\kappa(p)\cosh^{-1}(|a|^{-\frac{p}{\kappa(p)}}),
\end{equation}
the quantity
\begin{equation}\label{eq:LR-family-U}
v(\sigma,p)=\frac{\kappa(p)}{\sigma}\sinh^{-1}\left(|a|^{\frac{p}{\kappa(p)}}\sinh\left(\frac{\sigma}{\kappa(p)}\right)\right),
\end{equation}
is an LR-velocity for $U_\Phi$. In particular, there exists a constant $C(\sigma,p)<\infty$ such that
\begin{equation}\label{eq:LR-bound-U}
\|U^n_\Phi(j,k)\|\leq C(\sigma,p)\ e^{-\frac{\sigma}{p}(|j-k|-v(\sigma,p)|n|)},
\end{equation}
for all $j,k,n\in\Z$.
\end{theorem}

Theorem~\ref{thm:LR-bound} is proven in Sections~\ref{sec:Proof-odd} and \ref{sec:Proof-even}.

\begin{corollary} \label{cor:vLReqvAsy} 
The infimal LR-velocity and asymptotic velocity of $U_\Phi$ are equal, i.e.\
\begin{equation}
v_{\LR}^{\inf}(U_\Phi) = v_{\Asy}(U_\Phi) = |a|^{p/\kappa(p)}.
\end{equation}
\end{corollary}

\begin{proof} 
The extremal cases $a=0$ and $|a|=1$ are elementary and can be treated by directly following the path of a standard basis vector $|n, \pm\rangle$ under the walk $U_\Phi$. 

Taking $\sigma\to 0^+$ for $0<|a|<1$ gives
\begin{equation} \label{sigmatozero}
\lim_{\sigma\to 0^+}v(\sigma,p)=|a|^{\frac{p}{\kappa(p)}}=v_\Asy(U_\Phi).
\end{equation}
Combining this limit with \eqref{vAsyvLR}, we conclude that for all $a\in\overline{\mathbb{D}}$ 
\begin{equation}
v_{\LR}^{\inf}(U_\Phi)=v_{\Asy}(U_\Phi).
\end{equation}
\end{proof}

The one-parameter family of LR-velocities in \eqref{eq:LR-family-U} associates each admissible value of $\sigma$ with a fixed spatial decay rate. It therefore makes explicit the trade-off between the propagation velocity and the spatial decay of the bound. In particular, $\sigma$ is not merely an auxiliary optimization parameter: it controls how rapidly the bound decays outside the effective light cone. The optimal velocity is recovered in the limit $\sigma\downarrow 0$, whereas the full $\sigma$-dependent expression provides quantitative estimates at finite distances and times.

This trade-off has a direct probabilistic consequence. For all $j,n\in\Z$, define the position probability distribution after $n$ steps for a walker starting from $|j,s\rangle$, $s\in \{\pm\}$
\begin{equation}
p_{j,n}(k)=\sum_{s'\in\{+,-\}}|\langle k,s'| U_\Phi^n|j,s\rangle|^2, \quad k\in\Z.
\end{equation}

The one-parameter family of LR-velocities in Theorem \ref{thm:LR-bound} yields as a direct consequence the following exponential tail estimate for the total probability that after $n$ time steps a walker initially in $|j,s\rangle$ lies outside the Lieb--Robinson light cone of radius $v_\epsilon|n|$:
\begin{corollary}\label{coro:tail}
Assume $0<|a|<1$. For sufficiently small $\epsilon>0$, let $v_\epsilon=v^{\inf}_{\LR}(U_\Phi)+\epsilon$. There exists $\tilde{C}_\epsilon<\infty$ such that
\begin{equation}\label{eq:tail}
\sum_{k:\ |j-k|>v_\epsilon|n|}p_{j,n}(k)\leq \tilde{C}_\epsilon e^{-\gamma_\epsilon|n|}
\end{equation}
for all $j,n\in\Z$, where
\begin{equation}\label{eq:gamma-asymptotics}
\gamma_\epsilon
=\frac{4\kappa(p)}{3p\sqrt{A_p}}\epsilon^{3/2}+\mathcal{O}(\epsilon^{5/2})
\qquad \text{as }\epsilon\to0^+, \ \ \text{ and }\ A_p:=\frac{1}{2}\left(|a|^{\frac{p}{\kappa(p)}} - |a|^{\frac{3p}{\kappa(p)}}\right).
\end{equation}
Moreover, $\gamma_\epsilon$ is sharp in the sense that it is the largest decay exponent obtained by optimizing the family of bounds in \eqref{Tail:LR}.
\end{corollary}
The proof of Corollary \ref{coro:tail} is included in Appendix \ref{sec:lightcone}.

\begin{remark}
For general translation-invariant quantum dynamics, exponential decay outside the propagation cone and the characteristic $\epsilon^{3/2}$ behavior near its boundary were established via a large deviation principle in \cite{CJWW25}. In fact, both the uniform LR-velocity bound as well as the asymptotic bound in \cite{CJWW25} can be traced back to bounding the spectral radius of the tilted walk operator $W_{-i\sigma}:=e^{\sigma\cdot Q}We^{-\sigma\cdot Q}$. Optimizing this spectral radius over the tilts $\sigma$ gives the rate function in \cite{CJWW25}. At the same time, dividing this spectral radius by $\sigma$ yields the velocity in Theorem \ref{thm:LR-bound}.
\end{remark}

The proof of Theorem \ref{thm:LR-bound} is divided into Sections \ref{sec:Proof-odd} and \ref{sec:Proof-even}, which treat the odd- and even-period cases, respectively. The overall strategy as outlined at the beginning of the next section is the same. The odd-period case is simpler, as suggested by the form of \eqref{ExplOmega}. The even-period case requires sufficiently many distinct details that we repeat most of the argument in Section \ref{sec:Proof-even}.

\subsection{Electric split-step walks}\label{sec:W}
As the name suggests, the split-step quantum walk on $\mathcal{H}$ is defined by splitting the the operator implementing the steps by inserting an additional coin between the shift to the left and to the right:
\begin{equation}\label{def:W}
W:=S_+ C S_- C,
\end{equation}
where $S_\pm$ act on the canonical basis vectors $|n,\sigma\rangle$, $n\in\Z$ and $\sigma\in\{+,-\}$, according to
\begin{equation}\label{def:SPM}
\begin{aligned}
S_+|n,+\rangle &= |n+1,+\rangle, \quad &S_+|n,-\rangle &= |n,-\rangle,\\
S_-|n,+\rangle &= |n,+\rangle, \quad &S_-|n,-\rangle &= |n-1,-\rangle,
\end{aligned}
\end{equation}
and $C$ is a coin operator as in Section \ref{sec:U}. We are again especially interested in the translation invariant case, when the local coins are defined as in \eqref{def:C}. In analogy with the electric shift-coin walk, consider the local field
\begin{equation} \label{def:F-tilde}
\tilde{F}_\Phi := \left(\idty_{\Z} \otimes \begin{bmatrix} 1 & 0 \\ 0 & e^{2\pi i \Phi} \end{bmatrix} \right) F_{2\Phi} = e^{i\pi\Phi}e^{-i\pi \Phi(\idty_{\Z} \otimes \sigma_3)} F_{2\Phi},
\end{equation}
where $F_{\Phi}$ is as in \eqref{def:U-Phi} and $\sigma_3$ is the third Pauli matrix. We then define
\begin{equation}\label{def:W-Phi}
W_\Phi:=\tilde{F}_\Phi W.
\end{equation}
It was shown in \cite{ARS2023-CMP,ARCSW25} that the asymptotic velocity of this electric split-step walk is exponentially suppressed for rational fields. 

The parity of the numerator $q$ determines whether the reduced denominator of $\Phi/2$ is $2p$ or $p$. To record this distinction, let
\begin{equation}\label{def:kappa-tilde}
\tilde{\kappa}(q):=
\begin{cases}
2 & q \text{ is odd},\\
1 & q \text{ is even}.
\end{cases}
\end{equation}
The following Lieb--Robinson bound is obtained by transferring Theorem \ref{thm:LR-bound} through the sieving relation reviewed below.
\begin{corollary}\label{coro:W-LR}
Let $W_\Phi$ be the electric split-step quantum walk \eqref{def:W-Phi} with rational field $\Phi=q/p$, where $q$ and $p$ are relatively prime. Assume that the coin parameter $a$ in \eqref{def:C} satisfies $0<|a|<1$.
Then, with $\tilde{\kappa}(q)$ as in \eqref{def:kappa-tilde}, for every $\sigma$ satisfying
\begin{equation}\label{eq:sigma-range-W}
0<\sigma<\tilde{\kappa}(q)\cosh^{-1}(|a|^{-p}),
\end{equation}
the quantity
\begin{equation}\label{eq:LR-family-W}
\tilde{v}(\sigma,\Phi)=\frac{\tilde{\kappa}(q)}{\sigma}\sinh^{-1}\left(|a|^{p}\sinh\left(\frac{\sigma}{\tilde{\kappa}(q)}\right)\right),
\end{equation}
is an LR-velocity for $W_\Phi$. In particular,
\begin{equation}\label{eq:LR-bound-W}
\|W^n_\Phi(j,k)\|\leq C\left(\sigma,\tilde{\kappa}(q)p\right)\ e^{-\frac{\sigma}{p}(|j-k|-\tilde{v}(\sigma,\Phi)|n|)},
\end{equation}
for all $j,k,n\in\Z$. Here the constant $C$ is the same as that in Theorem \ref{thm:LR-bound}.
\end{corollary}
Before giving the proof, we record an immediate consequence of the corollary.
For $0<|a|<1$, both cases in \eqref{def:kappa-tilde} give
\begin{equation}
v_{\LR}^{\inf}(W_\Phi)\leq \lim_{\sigma\to 0^+}\tilde{v}(\sigma,\Phi)=|a|^p.
\end{equation}
The reverse inequality follows from \eqref{vAsyvLR} and \cite[Theorem 2.1(a)]{ARCSW25}. Thus
\begin{equation}
v_{\LR}^{\inf}(W_\Phi)= v_{\Asy}(W_\Phi) = |a|^p.
\end{equation}
For $a=0$ and $|a|=1$, direct inspection gives $v_{\LR}^{\inf}(W_\Phi)=0$ and $v_{\LR}^{\inf}(W_\Phi)=1$, respectively. Consequently, the infimal LR-velocity agrees with the asymptotic velocity for all $0\leq|a|\leq1$. Although this infimal LR-velocity is independent of the parity of $q$, the LR-velocity at a fixed decay rate retains the parity dependence through $\tilde{\kappa}(q)$.

\begin{proof}[Proof of Corollary \ref{coro:W-LR}] The sieving result in \cite[Section 3.2]{ARCSW25} states that the square of a shift-coin walk decomposes, with respect to the even--odd decomposition of the Hilbert space, into two split-step walks. Concretely, consider $\mathcal{H}_e = \ell^2(2\Z;\C^2)$ and $\mathcal{H}_o = \ell^2(2\Z+1;\C^2)$ as subspaces of $\mathcal{H} = \ell^2(\Z;\C^2)$. Clearly,
\begin{equation} \label{directsum}
\mathcal{H} = \mathcal{H}_e \oplus \mathcal{H}_o = \ell^2(2\Z;\C^2) \oplus \ell^2(2\Z+1;\C^2).
\end{equation}
Introduce partial isometries $E:\mathcal{H} \to \mathcal{H}$ and $O:\mathcal{H} \to \mathcal{H}$ given by
\begin{equation}
E|n,\pm \rangle = |2n, \pm \rangle, \quad O|n,\pm \rangle = |2n+1, \pm\rangle,
\end{equation}
and note that $E^* E = O^* O = \idty$, while $EE^*$ and $OO^*$ are the orthogonal projections onto $\mathcal{R}(E) = \mathcal{H}_e$ and $\mathcal{R}(O) = \mathcal{H}_o$, respectively. The following identity is a special case of \cite[Theorem 3.3]{ARCSW25}: for $U$ and $W$ defined in \eqref{def:U} and \eqref{def:W}, respectively,
\begin{equation}
U^2 = E W E^* \oplus O W O^*,
\end{equation}
with respect to the decomposition \eqref{directsum}.

To relate the electric shift-coin and split-step walks, we decompose $U_\Phi^2 = (F_\Phi U)^2$ with respect to \eqref{directsum}. Using the commutation relation between $U$ and $F_\Phi$ and then replacing $\Phi$ by $\Phi/2$ gives
\begin{equation} \label{UPhihalf1}
U_{\Phi/2}^2 = e^{-i\pi\Phi (\idty_{\Z} \otimes \sigma_3)} F_{\Phi} (E W E^* \oplus O W O^*).
\end{equation}
It remains to decompose $F_{\Phi}= e^{2\pi i \Phi Q}$. Under the identifications induced by $E$ and $O$, the position operator decomposes as $Q = (2Q) \oplus (2Q+\idty)$. We therefore obtain
\begin{equation} \label{UPhihalf2}
U_{\Phi/2}^2 = (e^{-i\pi \Phi} E W_\Phi E^*) \oplus (e^{i\pi \Phi} O W_\Phi O^*).
\end{equation}
Here we have also used the invariance of $\mathcal{H}_e$ and $\mathcal{H}_o$ under $\tilde{F}_\Phi$.
Iterating \eqref{UPhihalf2} gives
\begin{equation} \label{UPhihalf3}
U_{\Phi/2}^{2n} = (e^{-i\pi n \Phi} E W_\Phi^n E^*) \oplus (e^{i\pi n \Phi} O W_\Phi^n O^*),
\end{equation}
for all $n\in \N$; see \cite[Corollary 3.4]{ARCSW25}. In fact, \eqref{UPhihalf3} holds for all $n\in \Z$. The case $n=0$ is immediate, while the assertion for negative $n$ follows by taking adjoints in \eqref{UPhihalf3} and using the unitarity of $U_{\Phi/2}$ and $W_\Phi$.

Together with the properties of $E$ and $E^*$, this identity shows that the even-indexed matrix elements of $U_{\Phi/2}^{2n}$ satisfy
\begin{equation}
e^{i \pi n \Phi} U_{\Phi/2}^{2n} (2j,2k) = W_\Phi^n(j,k),
\end{equation}
and, in particular,
\begin{equation} \label{WUrelation}
\|W_\Phi^n(j,k)\| = \|U_{\Phi/2}^{2n}(2j,2k)\|,
\end{equation}
for all $n\in \Z$. This is the key observation in the reduction of split-step walks to shift-coin walks.

\begin{description}[labelindent=0pt,leftmargin=0pt,itemsep=5pt,topsep=5pt]
\item[$q$ odd] In this case, $q$ and $2p$ are relatively prime, and we apply (the ``even'' case of) Theorem \ref{thm:LR-bound} to \eqref{WUrelation} with $\Phi/2 = q/(2p)$. More precisely, if $0<\sigma <2\cosh^{-1}(|a|^{-p})$, then \eqref{eq:LR-family-U} and \eqref{eq:LR-bound-U} give
\begin{eqnarray} \label{Case1estimate}
\|W_\Phi^{n}(j,k)\| & = & \|U_{\frac{q}{2p}}^{2n}(2j,2k)\| \nonumber \\
& \leq & C(\sigma,2p) \exp \left[ -\frac{\sigma}{2p} \left( |2j-2k|-2 v(\sigma,2p) |n| \right) \right] \nonumber \\
& = & C(\sigma,2p) \exp \left[ -\frac{\sigma}{p} \left( |j-k|- v(\sigma,2p) |n| \right) \right],
\end{eqnarray}
for all $j,k,n\in\Z$. Thus the corresponding family of LR-velocities for $W_\Phi$ is
\begin{equation}\label{eq:W-Phi-q-odd}
v(\sigma,2p)=\frac{2}{\sigma}\sinh^{-1}\left(|a|^{p}\sinh(\sigma/2)\right).
\end{equation}
 
\item[$q$ even] In this case, $p$ is odd and $q/2$ and $p$ are relatively prime. For $0<\sigma < \cosh^{-1}(|a|^{-p})$, we argue as in the odd-$q$ case and use \eqref{ngeneral} (thus applying the ``odd'' case of Theorem~\ref{thm:LR-bound}) to obtain
\begin{equation}\label{Case2estimate}
\|W_\Phi^{n}(j,k)\| \leq C(\sigma,p) \exp \left[ - \frac{2\sigma}{p} \left( |j-k|- v(\sigma,p) |n| \right) \right],
\end{equation}
where here (with odd $p$ in \eqref{eq:LR-family-U}), 
\begin{equation}\label{eq:W-Phi-q-even}
v(\sigma,p)=\frac{1}{\sigma}\sinh^{-1}\left(|a|^p\sinh\sigma\right)
\end{equation}
\end{description}

The two cases \eqref{Case1estimate} and \eqref{Case2estimate}, and  the LR-velocities in \eqref{eq:W-Phi-q-odd} and \eqref{eq:W-Phi-q-even} are written in the compact way \eqref{eq:LR-family-W}. This completes the proof of Corollary \ref{coro:W-LR}.
\end{proof}

\section{Proof of the LR-velocity bound: the odd period}\label{sec:Proof-odd}
We start by providing an outline of the proof (for odd or even period).

The central calculation will start with the application of \eqref{Bloch} to $W=U_\Phi^p$, i.e.\
\begin{equation} \label{UBloch}
U_\Phi^p(j,k) = \frac{1}{2\pi} \int_0^{2\pi} \widehat{U_\Phi^p}(e^{i\theta}) e^{-i(j-k)\theta}\,d\theta.
\end{equation}
As mentioned earlier, the Floquet matrix $\widehat{U_\Phi^p}(e^{i\theta})$ can be analytically extended to $z\in\C\backslash\{0\}$. We begin with showing that the two eigenvalues of this extension $\widehat{U_\Phi^p}(z)$ are not degenerate within some annulus $1/\rho < |z|< \rho$ (Lemma~\ref{lem:simplicity-max}), and we establish a lower bound on the distance of these eigenvalues (Lemma~\ref{lem:odd-gap}). Then we will apply \eqref{UBloch} to $(\widehat{U_\Phi^p}(z))^n$, rewrite this integral as a contour integral over the complex unit circle and deform the contour to suitable circles within the annulus.  As $\widehat{U_\Phi^p}(e^{i\theta})$ is unitary with determinant one, the eigenvalues can be written as $\lambda_{\pm}(e^{i\theta},a) = e^{\pm i \omega(e^{i\theta},a)}$ with real phase $\omega(e^{i\theta},a)$. Under extension of $\widehat{U_\Phi^p}(z)$ to non-unit circles, this phase will pick up a non-trivial imaginary part as $\widehat{U_\Phi^p}(z)$ may be non-unitary for $|z|\not= 1$, which will determine the behavior of \eqref{UBloch}. An upper bound on the modulus of the imaginary part of the phase will be provided in Lemma~\ref{lem:gap}. We will then fully diagonalize $\widehat{U_\Phi^p}(z)$ as
\begin{equation} \widehat{U_\Phi^p}(z) = e^{i\omega(z,a)} P_+(z) + e^{-i\omega(z,a)} P_-(z)
\end{equation} 
with the (non orthogonal) spectral projections $P_{\pm}(z)$ \cite[Chapter 1]{Kato} . We will bound the norm of the projections $P_\pm(z)$ with the help of the elementary formula 
\begin{equation} \label{boundnormhatU}
P_{\pm}(z) = \frac{\widehat{U_\Phi^p}(z)-\lambda_{\mp}(z,a)\idty}{\lambda_{\pm}(z,a)-\lambda_{\mp}(z,a)}.
\end{equation}

 This bound will use the eigenvalue gap established in Lemma~\ref{lem:odd-gap}. A final limiting process in the radius of the annuli will yield $v_{\LR}^{\inf}(U_\Phi^p) \leq p|a|^p$. We then conclude the proof by applying Lemma~\ref{scaling} and \eqref{vAsyvLR} and the fact that the asymptotic velocity $v_{\Asy}(U_\Phi)$ is $|a|^p$ and $|a|^{p/2}$ in the case of odd $p$ and even $p$, respectively \cite{ARCSW25}.
 
\subsection{Simplicity of the spectrum of the extended Floquet matrix} \label{simplicity}
For $\rho>1$, define the annulus 
\begin{equation}\label{def:A}
\mathcal{A}(\rho):=\{z\in\C:\ \rho^{-1}<|z|< \rho\}.
\end{equation}
\begin{lemma}\label{lem:simplicity-max}
Let  $p\geq 1$ be odd, $|a|\in (0,1)$, and 
\begin{equation}\label{def:rho0-odd}
\rho_{o}(|a|):=\frac{\left(1+\sqrt{1-|a|^{2p}}\right)^{1/p}}{|a|}.
\end{equation}
Then, $\widehat{U_\Phi^p}(z)$ has simple spectrum on $\mathcal{A}(\rho_{o}(|a|))$. In fact, $\mathcal{A}(\rho_{o}(|a|))$ is the largest possible annulus on which $\widehat{U_\Phi^p}(z)$ has simple spectrum.
\end{lemma}

\begin{proof}
Let us write $a=|a|e^{i\phi}$. Since by Lemma \ref{lem:trace} we have $\Tr(\widehat{U_\Phi^p}(z))=|a|^p(\tilde z^p+\tilde{z}^{-p})$ where $\tilde z=e^{i\phi}z$ and $\det \widehat{U_\Phi^p}(z)=1$, the characteristic polynomial of $\widehat{U_\Phi^p}(z)$ is 
\begin{equation}
\lambda^2-|a|^p(\tilde z^p+\tilde{z}^{-p})\lambda+1.
\end{equation}
Thus, the eigenvalues of  $\widehat{U_\Phi^p}(z)$ are degenerate if $|a|^p(\tilde{z}^p+\tilde{z}^{-p})=\pm 2$. 
First, let $|a|^p(\tilde{z}^p+\tilde{z}^{-p})=+ 2$. 
Multiplication by $\tilde{z}^p$ yields
\begin{equation}
\tilde{z}^{2p}-\frac{2}{|a|^p}+1=0 \qquad \Rightarrow \qquad \tilde{z}^{p}=\frac1{|a|^p}\pm\sqrt{\frac1{|a|^{2p}}-1}.
\end{equation}
Observe that these roots are reciprocals of each other and hence have moduli
\begin{equation} \label{modroots}
R_o^{\pm1}:=|a|^{-p}\pm\sqrt{|a|^{-2p}-1}=\frac{1\pm\sqrt{1-|a|^{2p}}}{|a|^p}.
\end{equation}
A similar argument shows that \eqref{modroots} holds in the case $|a|^p(\tilde{z}^p+\tilde{z}^{-p})=-2$. 
Hence, degeneracies occur for $|z|=R_o^{1/p}$ or $|z|=R_o^{-1/p}$.

Thus, if $R_o^{-1/p}<|z|<R_o^{1/p}$, then $\widehat{U_\Phi^p}(z)$ has simple spectrum.
\end{proof}

\begin{lemma}\label{lem:odd-gap}
Let  $p\geq 1$ be odd. If $a\in\C$ with $0<|a|<1$ and  
\begin{equation} \label{sigmabound1} 
0<\sigma < \cosh^{-1}(|a|^{-p}),
\end{equation}
then $\widehat{U_\Phi^p}(z)$ has simple spectrum for all $z \in \overline{\mathcal{A}(e^{\sigma/p})}$. Moreover, the distance between the eigenvalues $\lambda_+(z)$ and $\lambda_-(z)$ of $\widehat{U_\Phi^p}(z)$ has a uniform lower bound in $\overline{\mathcal{A}(e^{\sigma/p})}$, more precisely,
\begin{equation}\label{eq:odd-gap}
|\lambda_+(z)-\lambda_-(z)|\geq 2\sqrt{1-|a|^{2p}\cosh^2\sigma}>0
\end{equation}
for all $z\in\overline{\mathcal A(e^{\sigma/p})}$.
\end{lemma}
\begin{proof}
The simplicity of the spectrum on $\overline{\mathcal A(e^{\sigma/p})}$ follows from Lemma \ref{lem:simplicity-max} once we establish that
\begin{equation} \label{ToShow1}
e^{\sigma/p} < \rho_{o}(|a|)
\end{equation}
with $\rho_{o}(|a|)$ as in \eqref{def:rho0-odd}.
Note that $\rho_{o}(t)$ is a strictly decreasing function of $t\in (0,1)$ with
\begin{equation} \label{limrho} 
\lim_{t\to 0^+} \rho_{o}(t)=\infty \quad \mbox{and} \quad \lim_{t\to 1^-} \rho_{o}(t) = 1.
\end{equation}
A calculation shows that 
\begin{equation}
\rho_{o}^{-1}(e^{\sigma/p}) = (\cosh\sigma)^{-1/p},
\end{equation}
where the latter is a consequence of \eqref{sigmabound1}. As $\rho_{o}$ is decreasing with $\rho_{o}((0,1)) = (1,\infty)$, this implies \eqref{ToShow1}.

To prove \eqref{eq:odd-gap}, recall the characteristic equation $\lambda^2-\Tr(\widehat{U_\Phi^p}(z))\lambda+1= 0$. Thus, the two eigenvalues $\lambda_\pm(z)$ of $\widehat{U_\Phi^p}(z)$ satisfy
\begin{equation}\label{eq:pf:odd-gap-0}
|\lambda_+(z)-\lambda_-(z)|^2=\left|\Tr(\widehat{U_\Phi^p}(z))^2-4\right|\geq 4-\left|\Tr(\widehat{U_\Phi^p}(z))\right|^2.
\end{equation}
We also recall that $|\Tr(\widehat{U_\Phi^p}(z))| \leq |a|^p\left(|\tilde{z}|^p+|\tilde{z}|^{-p}\right)$. Hence, for any $z\in\overline{\mathcal A(e^{\sigma/p})}$
\begin{equation}
\left|\Tr(\widehat{U_\Phi^p}(z))\right|\leq 2 |a|^p\cosh\sigma.
\end{equation}
Plugging this into \eqref{eq:pf:odd-gap-0} (and taking the square root) yields \eqref{eq:odd-gap}.
\end{proof}

\subsection{Bounding the imaginary part of the dispersion relations} \label{ImPart}

For $z\in \C \setminus \{0\}$ with $|z|\not= 1$, $\widehat{U_\Phi^p}(z)$ may not be unitary, however its determinant is $1$. Thus its eigenvalues are reciprocals of each other and can therefore be written in the form $e^{\pm i\omega}$ for some $\omega \in \C_+$ with uniquely determined (positive) imaginary part. The next lemma provides a uniform bound on the latter for $z\in \overline{\mathcal{A}(e^{\sigma/p})}$, which ultimately will determine the LR-velocity.

\begin{lemma}\label{lem:gap}
Let $p\geq 1$ be odd, $0<|a|<1$, and $\sigma\in(0,1)$. If $z\in\overline{\mathcal A(e^{\sigma/p})}$ and $e^{i\omega}$ is an eigenvalue of $\widehat{U_\Phi^p}(z)$,	 then
\begin{equation}\label{odd-Im-bound}
|\Im \omega|\leq \sinh^{-1}\left( |a|^p\sinh\sigma\right).
\end{equation}
\end{lemma}

\begin{proof}
Recall
\begin{equation}
 \lambda_+(z)+\lambda_-(z) = \Tr(\widehat U_\Phi^p(z))= |a|^p(\tilde{z}^p+\tilde{z}^{-p}).
\end{equation}
Writing these eigenvalues as $e^{\pm i \omega}$, we obtain
\begin{equation}
\cos \omega=\frac{1}{2} |a|^p (\tilde{z}^p+\tilde{z}^{-p}).
\end{equation}
Since $|\tilde{z}|=|z|$ and $e^{-\sigma}\leq |z|^p \leq e^{\sigma}$, we may write
\begin{equation}
\tilde{z}^p=e^{u+i\theta}, \quad \text{ where }\quad -\sigma \leq u \leq \sigma,\: \theta \in [0,2\pi).
\end{equation}
Note that if $u=0$, then $\cos \omega$ is real and lies in $[-|a|^p,|a|^p]$,
so the eigenvalues lie on the unit circle and the desired bound is trivially satisfied.
Thus we may assume $u\neq 0$.

Then
\begin{equation}
\frac{1}{2}(\tilde{z}^p+\tilde{z}^{-p})= \cosh(u+i\theta) = \cosh u\cos\theta+i\sinh u\sin\theta.
\end{equation}
Thus, 
\begin{equation}
\cos \omega=|a|^p \cosh u\cos\theta+ i|a|^p \sinh u\sin\theta.
\end{equation}
Then observe that, the curve $\theta\mapsto\cos\omega$ traces an ellipse centered at the origin, with axes $|a|^p\cosh u$ and $|a|^p|\sinh u|$, as follows.
\begin{equation}\label{odd:ellipse}
\frac{\left(\Re (\cos \omega)\right)^2}{|a|^{2p}\cosh^2u}+\frac{\left(\Im(\cos \omega)\right)^2}{|a|^{2p}\sinh^2 u}=\cos^2\theta+\sin^2\theta=1.
\end{equation}
Since $-\sigma \leq u \leq \sigma$, 
\begin{equation}
\cosh^2 u=1+\sinh^2 u \quad \text{ and }\quad \sinh^2 u\leq \sinh^2\sigma.
\end{equation}
Moreover, use $|a|<1$,
\begin{equation}
|a|^{2p}\cosh^2 u\leq 1+|a|^{2p}\sinh^2\sigma.
\end{equation}
Thus, \eqref{odd:ellipse} gives
\begin{equation}\label{even:ellipse-2}
\frac{\left(\Re (\cos \omega)\right)^2}{1+|a|^{2p}\sinh^2\sigma}+\frac{\left(\Im(\cos \omega)\right)^2}{|a|^{2p}\sinh^2\sigma}\leq \cos^2\theta+\sin^2\theta=1.
\end{equation}
Let
\begin{equation}
\omega(z,a)=x+iy.
\end{equation}
Here we claim that \eqref{odd-Im-bound} is valid. 
Suppose, to the contrary, that $|y|>\sinh^{-1}(|a|^p\sinh \sigma)$, which is equivalent to
\begin{equation}\label{odd-bound-!!}
 \sinh^2 y>|a|^{2p}\sinh^2\sigma,
\end{equation}
 then substitute $\omega =x+iy$ in the LHS of \eqref{even:ellipse-2} and use
\begin{equation} \label{addformula}
\cos(x+iy)=\cos x\cosh y- i \sin x\sinh y,
\end{equation}
to obtain
\begin{equation}\label{even:ellipse:!!-1}
\cos^2 x\frac{\cosh^2 y}{1+|a|^{2p}\sinh^2 \sigma }+\sin^2 x\frac{\sinh^2 y}{|a|^{2p}\sinh^2 \sigma}.
\end{equation}
Then use the assumption \eqref{odd-bound-!!}
\begin{equation}
\cosh^2 y=1+\sinh^2 y>1+|a|^{2p}\sinh^2(\sigma)
\end{equation}
to see that \eqref{even:ellipse:!!-1} is bounded below by
\begin{equation}
\cos^2 x\frac{\cosh^2 y}{1+|a|^{2p}\sinh^2\sigma }+\sin^2 x\frac{\sinh^2 y}{|a|^{2p}\sinh^2 \sigma}  > \cos^2 x+\sin^2 x=1.
\end{equation}
This contradicts \eqref{even:ellipse-2}.

Hence,
\begin{equation}
\displaystyle |\Im \omega|\leq \sinh^{-1}\left(|a|^{p}\sinh \sigma \right).
\end{equation}
\end{proof}

\subsection{Deformation of a contour integral} \label{deformation}

Here we complete the proof of the odd case in Theorem~\ref{thm:LR-bound}. Other than bringing all earlier bounds together, at the heart lies the deformation of a contour integral.

Let $p\geq 1$ be odd and let $0<|a|<1$. Choose $\sigma=\sigma(a,p)$ as in \eqref{sigmabound1}.
Then, by Lemma \ref{lem:odd-gap}, the distance between the two eigenvalues of
$\widehat{U_\Phi^p}(z)$ satisfies
\begin{equation} \label{unifsep}
|\lambda_+(z)-\lambda_-(z)|
\geq
2\sqrt{1-|a|^{2p}\cosh^2\sigma}
>0
\end{equation}
uniformly in $z\in\overline{\mathcal A(e^{\sigma/p})}$.

By \eqref{Bloch} and \eqref{nPower}, the $(j,k)$-th $2$-by-$2$ block of $U_{\Phi}^{pn}$ is given by
\begin{equation}
U_{\Phi}^{pn}(j,k)
=
\frac{1}{2\pi}
\int_0^{2\pi}
\left(\widehat U_\Phi^p(e^{i\theta})\right)^n
e^{-i(j-k)\theta}\ d\theta \quad \mbox{for all } n\in \Z.
\end{equation}
With the change of variables $z=e^{i\theta}$, this becomes
\begin{equation}
U_\Phi^{pn}(j,k)=\frac{1}{2\pi i}\oint_{|z|=1}\left(\widehat{U_\Phi^p}(z)\right)^nz^{-(j-k)-1}\ dz.
\end{equation}

The matrix-valued integrand
\begin{equation}
\left(\widehat{U_\Phi^p}(z)\right)^n z^{-(j-k)-1}
\end{equation}
is analytic for $z\neq0$. Therefore the contour may be deformed within the annulus $\mathcal{A}(\rho_{o}(|a|))$ where $\rho_{o}(|a|)$ is given in \eqref{def:rho0-odd}. 

We first assume $j\geq k$ and deform the contour to the positively oriented circle $\{|z|=e^{\sigma/p}\} \subset\mathcal{A}(\rho_{o}(|a|))$,
\begin{equation} \label{Upnint}
U_\Phi^{pn}(j,k)=\frac{1}{2\pi i}\oint_{|z|=e^{\sigma/p}}\left(\widehat{U_\Phi^p}(z)\right)^nz^{-(j-k)-1}\ dz.
\end{equation}

We now bound the norm of the integrand on the deformed contour:

The spectral representation of $\widehat{U_\Phi^p}(z)$ is
\begin{equation} \label{specrep}
\widehat{U_\Phi^p}(z) = \lambda_+(z) P_+(z) + \lambda_-(z) P_-(z).
\end{equation}
The spectral projections $P_{\pm}(z)$ satisfy (see e.g., \cite[Chapter I.3.4, eqs.\ (3.27), (3.28)]{Kato})
\begin{equation} \label{sumI}
(P_{\pm}(z)) ^2 = P_{\pm}(z), \quad P_{+}(z)P_-(z) = P_-(z) P_+(z) =0,\ \text{ and } \ P_+(z) + P_-(z) = \idty.
\end{equation}
 The eigen nilpotents vanish as $\lambda_{\pm}(z)$ are distinct. It follows that
\begin{equation} \label{specrepn}
\left(\widehat{U_\Phi^{p}}(z)\right)^n = (\lambda_+(z))^n P_+(z) + (\lambda_-(z))^n P_-(z).
\end{equation}

The norm of the spectral projections is uniformly bounded on the deformed contour. Indeed, by \eqref{specrep} and \eqref{sumI},
\begin{equation}
P_+(z)=\frac{\widehat{U_\Phi^p}(z)-\lambda_-(z)\idty}{\lambda_+(z)-\lambda_-(z)},\qquad P_-(z)=\frac{\widehat{U_\Phi^p}(z)-\lambda_+(z)\idty}{\lambda_-(z)-\lambda_+(z)}.
\end{equation}
We have $|\lambda_\pm(z)|\leq \|\widehat{U_\Phi^p}(z)\|$ and thus 
\begin{equation}
\|P_{\pm}(z)\|\leq \frac{2\|\widehat{U_\Phi^p}(z)\|}{|\lambda_+(z)-\lambda_-(z)|}
\end{equation}
which is uniformly bounded on $\{|z|=e^{\sigma/p}\}$ due to continuity of $\widehat{U_\Phi^p}(z)$ and \eqref{unifsep}.
Thus there exists $C(\sigma,p)<\infty$ such that
\begin{equation} \label{Ppmbound}
\sup_{|z|=e^{\pm\sigma/p}}
\left(\|P_+(z)\|+\|P_-(z)\|\right)
\leq C(\sigma,p).
\end{equation}

Since $\det \widehat{U_\Phi^p}(z)=1$, the two eigenvalues are reciprocal. Thus, pointwise on the deformed contour, we may write
\begin{equation}
\lambda_+(z) = e^{i\omega(z,a)}, \quad \lambda_-(z) = e^{-i\omega(z,a)}
\end{equation}
for some $\omega(z,a) \in \C$.
Then \eqref{specrepn} becomes
\begin{equation} \label{diagonalization}
\left(\widehat{U_\Phi^p}(z)\right)^n = e^{in\omega(z,a)}P_+(z) + e^{-in\omega(z,a)}P_-(z).
\end{equation}

By Lemma \ref{lem:gap}, for all $z\in\overline{\mathcal A(e^{\sigma/p})}$,
\begin{equation}
|\Im\omega(z,a)| \leq Y_o(\sigma,p):= \sinh^{-1}\left(|a|^p\sinh\sigma\right).
\end{equation}
Therefore, on the deformed contours,
\begin{equation}
\left\|\left(\widehat{U_\Phi^p}(z)\right)^n\right\| \leq C(\sigma,p)e^{n Y_o(\sigma,p)}.
\end{equation}
Now \eqref{Upnint} gives the bound
\begin{equation}  \label{normUpnbound}
\|U_\Phi^{pn}(j,k)\| \leq \frac{1}{2\pi}\oint_{|z|=e^{\sigma/p}} \left\|\left(\widehat{U_\Phi^p}(z)\right)^n\right\| |z|^{-(j-k)-1}\ |dz|.
\end{equation}
Since $|z|=e^{\sigma/p}$ and $|dz|=e^{\sigma/p}d\theta$ on this contour, we obtain
\begin{equation}
\|U_\Phi^{pn}(j,k)\| \leq C(\sigma,p) e^{n Y_o(\sigma,p)} e^{-(j-k)\sigma/p}.
\end{equation}

Similarly, if $j< k$, we deform the contour to $\{|z|=e^{-\sigma/p}\}$ and obtain
\begin{equation}
\|U_\Phi^{pn}(j,k)\| \leq C(\sigma,p)e^{n Y_o(\sigma,p)}e^{-(k-j)\sigma/p}.
\end{equation}
Therefore, for all $j,k$,
\begin{equation} \label{ngreater0}
\|U_\Phi^{pn}(j,k)\| \leq C(\sigma,p)\exp\left[-\frac{\sigma}{p}\left(|j-k| -\frac{p}{\sigma} Y_o(\sigma,p)n\right)\right].
\end{equation}

We finally comment on how to extend this to arbitrary $n\in \Z$. The case $n=0$ is trivial. Let $n<0$. For this observe that, by \eqref{Floquetinv}, $U_\Phi^{-p}$ is also a translation invariant quantum walk with
\begin{equation} 
\widehat{U_\Phi^{-p}}(e^{i\theta}) = \left( \widehat{U_\Phi^p} (e^{i\theta}) \right)^{-1}.
\end{equation}
This allows to carry out virtually the same proof as above with the roles of $\lambda_-(z)$ and $\lambda_+(z)$ interchanged (as they are reciprocals). Ultimately, one gets
\begin{equation} \label{nlesseq0}
\|(U_\Phi^{-p})^{|n|} (j,k) \| \leq C(\sigma,p)\exp\left[-\frac{\sigma}{p}\left(|j-k| -\frac{p}{\sigma} Y_o(\sigma,p)|n|\right)\right].
\end{equation}

Combining \eqref{ngreater0} and \eqref{nlesseq0} we find that 
\begin{equation} \label{ngeneral}
\|U_\Phi^{pn}(j,k)\| \leq C(\sigma,p)\exp\left[-\frac{\sigma}{p}\left(|j-k| -\frac{p}{\sigma} Y_o(\sigma,p)|n|\right)\right].
\end{equation}
for all $j, k, n \in \Z$. Thus, for every $\sigma>0$ as in \eqref{sigmabound1},
\begin{equation}
\frac{p}{\sigma} Y_o(\sigma,p) = \frac{p}{\sigma}\sinh^{-1}\left(|a|^p\sinh\sigma\right).
\end{equation}
is a LR-velocity bound for $U_\Phi^p$. Lemma \ref{scaling} and its proof (to assure that the constant $c$ in Lemma~\ref{scaling} isn't changed when scaling from $U_\Phi^{pn}$ to $U_\Phi^n$) completes the 
proof of Theorem \ref{thm:LR-bound} for odd period $p$.

\section{Proof of the LR-velocity bound: the even period} \label{sec:Proof-even}

In this section we give the proof of Theorem~\ref{thm:LR-bound} for the case of even $p$. We give a widely self-contained presentation of this case because of the significantly more involved details.

\subsection{Simplicity of the spectrum of the extended Floquet matrix}

Assume that $p\geq 2$ is even, $a=e^{i\phi}|a|$, $0<|a|<1$.
The even case of Lemma \ref{lem:trace} gives, with $\tilde{z}:=e^{i\phi}z$,
\begin{equation} \label{eq:trace-even}
\Tr(\widehat{U_\Phi^p}(z))=-|a|^p\left(\widetilde z^p+\widetilde z^{-p}\right)+2s_p(|a|^p-1), \text{ where }s_p:=(-1)^{p/2}.
\end{equation}
Recall the definition of an annulus $\mathcal{A}(\rho)$ in \eqref{def:A}.

We show that the eigenvalues of the extended Floquet matrix are simple on suitable annuli, with the exception of some singular point on the unit circle.

\begin{lemma}\label{lem:simplicity-even}
Let $p\geq 2$ be even. For a given $a=e^{i\phi}|a|\in\C$ with $|a|\in(0,1)$, the Floquet matrix $\widehat{U_\Phi^p}(z)$ has simple spectrum for
\begin{equation}
z\in \mathcal A(\rho_{0,e}(|a|) )\setminus \mathcal Z_p(a),
\end{equation}
where
\begin{equation} \label{def:Z}
\rho_{0,e}(t) = \left(\frac{1+\sqrt{1-t^p}}{1-\sqrt{1-t^p}}\right)^{1/p} \quad \text{ and } \quad \mathcal Z_p (a):=\left\{z\in\mathbb C:\ (e^{i\phi} z)^p=(-1)^{p/2}\right\}.
\end{equation}
Moreover, $\widehat{U_\Phi^p}(z)$ is degenerate at every point of $\mathcal Z_p(a)$. \footnote{Note that $|\mathcal Z_p(a)|=p$}
\end{lemma}

\begin{proof}[Proof of Lemma \ref{lem:simplicity-even}]
As for $p$ odd, we have that the spectrum is simple if and only if
\begin{equation}
\Tr(\widehat{U_\Phi^p}(z))\neq \pm 2.
\end{equation}

Let $\eta:=\widetilde z^p$.
Using \eqref{eq:trace-even}, degeneracy occurs precisely when 
\begin{equation}
-|a|^p(\eta+\eta^{-1})+2s_p(|a|^p-1)=2\varepsilon, 
\end{equation}
where $\epsilon \in \{+1,-1\}$, and recall that $s_p = (-1)^{p/2}$.
Equivalently,
\begin{equation}\label{odd:eq:main-ref}
\eta+\eta^{-1}=2 \left(\frac{s_p(|a|^p-1)-\varepsilon}{|a|^p}\right).
\end{equation}

Here there are exactly two possibilities, namely, $\varepsilon=s_p$ or $\varepsilon=-s_p$.\\

\noindent \underline{Case 1: $\epsilon = -s_p$:} Then \eqref{odd:eq:main-ref} reads as
\begin{equation}
\eta+\eta^{-1}=2s_p \quad\Rightarrow\quad \eta=s_p=(-1)^{p/2}.
\end{equation}
Therefore degeneracy occurs whenever
\begin{equation}\label{odd:deg-1}
\widetilde z^p=(-1)^{p/2}.
\end{equation}
These are $p$ points on the unit circle.\\

\noindent \underline{Case 2: $\varepsilon=s_p$:} Then \eqref{odd:eq:main-ref} reduces to
\begin{equation}
\eta+\eta^{-1}
=-2s_p c \text{\quad where \quad}c:=\frac{2-|a|^p}{|a|^p}>1.
\end{equation}
The solutions are
\begin{equation}
\eta=-s_p\left(c\pm\sqrt{c^2-1}\right).
\end{equation}
The moduli of these two roots are (they are reciprocals)
\begin{equation}
R_{e}^{\pm 1}=c\pm\sqrt{c^2-1}=\frac{2-|a|^p\pm2\sqrt{1-|a|^p}}{|a|^p}=\frac{1\pm \sqrt{1-|a|^p}}{1\pm \sqrt{1-|a|^p}}.
\end{equation}
So, beside the degeneracies on the unit circle in \eqref{odd:deg-1}, the remaining degeneracies occur when $|z|=R_e^{\pm 1/p}$.

This proves the claim.
\end{proof}

Next, we quantify the previous lemma by establishing a lower bound on the distance of the eigenvalues, uniform on an annulus with removed neighborhoods of the singular points.

\begin{lemma}\label{lem:even-gap-away}
Let $p\geq 2$ be even. If $a\in \C$ with $0<|a|<1$ and
\begin{equation} \label{sigmabound}
0 < \sigma < 2 \cosh^{-1}(|a|^{-p/2}), 
\end{equation}
then $\widehat{U_\Phi^p}(z)$ has simple spectrum for all $z\in \overline{\mathcal{A}(e^{\sigma/p})} \setminus \mathcal{Z}_p(a)$. 

Moreover, For $\delta>0$, define
\begin{equation}
K_{\sigma,\delta}(a):=\left\{z\in\overline{\mathcal A(e^{\sigma/p})}:\ |\widetilde z^p-s_p|\geq \delta\right\}.
\end{equation}
Then the gap between the two eigenvalues $\lambda_+(z)$ and $\lambda_-(z)$ of $\widehat{U_\Phi^p}(z)$ is uniformly bounded from below on $K_{\sigma,\delta}(a)$. More precisely,
\begin{equation}\label{eq:even-gap-away}
|\lambda_+(z)-\lambda_-(z)|\geq 2\delta |a|^{p/2}e^{-\sigma/2}\sqrt{1-|a|^p\cosh^2(\sigma/2)}
\end{equation}
for all $z\in K_{\sigma,\delta}(a)$.
\end{lemma}

\begin{proof}
We first the simplicity of the spectrum on $\overline{\mathcal{A}(e^{\sigma/p})}\setminus\mathcal{Z}_p(a)$. By Lemma~\ref{lem:simplicity-even} we have to show that
\begin{equation} \label{ToShow}
e^{\sigma/p} < \rho_{e}(|a|).
\end{equation}
Note that $\rho_{e}(t)$ is a strictly decreasing function in $t\in(0,1)$ with
\begin{equation}
\lim_{t\to 0^+} \rho_{e}(t) \to \infty \quad \text{ and }\quad \lim_{t\to 1^-} \rho_{e}(t) = 1.
\end{equation}
A calculation shows that
\begin{equation}
\rho_{e}^{-1}(e^{\sigma/p}) = \left(\cosh(\sigma/2)\right)^{-2/p} > |a|,
\end{equation} 
where the latter follows from \eqref{sigmabound}. As $\rho_{e}$ is decreasing with $\rho_{0,a}((0,1)) = (1,\infty)$, this implies \eqref{ToShow}.

In the following, we prove the uniform bound on the gap of eigenvalues \eqref{eq:even-gap-away}. Recall the trace formula \eqref{eq:trace-even},
\begin{equation}
\Tr(\widehat U_p^p(z))
=-|a|^p(\widetilde z^p+\widetilde z^{-p})+2s_p(|a|^p-1).
\end{equation}
Since $\det\widehat{U_\Phi^p}(z)=1$, the two eigenvalues satisfy (use $s_p^2=1$)
\begin{equation}\label{eq:even:+-}
(\lambda_+(z)-\lambda_-(z))^2=\left(\Tr(\widehat{U_\Phi^p}(z))\right)^2-4
=\left(\Tr(\widehat{U_\Phi^p}(z))+2s_p\right)
\left(\Tr(\widehat{U_\Phi^p}(z))-2s_p\right).
\end{equation}
Next, we bound the two factors in \eqref{eq:even:+-}.

First,
\begin{equation}
\Tr(\widehat{U_\Phi^p}(z))+2s_p
= -|a|^p(\widetilde z^p+\widetilde z^{-p}-2s_p)
=-|a|^p\frac{(\widetilde z^p-s_p)^2}{\widetilde z^p}.
\end{equation}
Hence, for $z\in K_{\sigma,\delta}(a)$,
\begin{equation}\label{eq:even:+}
\left|\Tr(\widehat{U_\Phi^p}(z))+2s_p\right| =|a|^p\frac{|\widetilde z^p-s_p|^2}{|z|^p}\geq |a|^p\delta^2 e^{-\sigma},
\end{equation}
because $|z|^p\leq e^\sigma$ on $\overline{\mathcal A(e^{\sigma/p})}$.

Second,
\begin{equation}
\Tr(\widehat{U_\Phi^p}(z))-2s_p =-|a|^p(\widetilde z^p+\widetilde z^{-p})-2s_p(2-|a|^p).
\end{equation}
Therefore, 
\begin{equation}
\left|\Tr(\widehat{U_\Phi^p}(z))-2s_p\right| \geq 2(2-|a|^p)-|a|^p|\widetilde z^p+\widetilde z^{-p}| \geq 2(2-|a|^p)-2 |a|^p\cosh\sigma.
\end{equation}
Using $1+\cosh\sigma=2\cosh^2(\sigma/2)$, this becomes
\begin{equation}\label{eq:even:-}
\left|\Tr(\widehat{U_\Phi^p}(z))-2s_p\right| \geq 4\left(1-|a|^p\cosh^2(\sigma/2)\right)>0,
\end{equation}
where the strict positivity follows from \eqref{sigmabound}

Use \eqref{eq:even:+} and \eqref{eq:even:-} in \eqref{eq:even:+-} gives
\begin{equation}
|\lambda_+(z)-\lambda_-(z)|^2
\geq
4|a|^p\delta^2 e^{-\sigma}
\left(1-|a|^p\cosh^2(\sigma/2)\right).
\end{equation}
which proves the lemma.
\end{proof}

\subsection{Bounding the imaginary part of the dispersion relation}
Now we arrive at the key technical lemma, bounding the imaginary part of the dispersion relation of the extension of the Floquet matrix. The main insight behind its proof is to recognize that a certain expression describes an ellipse.

\begin{lemma}\label{lem:even-Im-bound}
Let $p\geq 2$ be even, $0<|a|<1$, and $\sigma >0$. Suppose that $z\in\overline{\mathcal A(e^{\sigma/p})}$ and $e^{i\omega}$ is any eigenvalue of $\widehat{U_\Phi^p}(z)$. Then
\begin{equation}\label{even-Im-bound}
|\Im \omega| \leq 2\sinh^{-1}\left(|a|^{p/2}\sinh(\sigma/2)\right).
\end{equation}
\end{lemma}

\begin{proof}
For even $p$, we have by \eqref{eq:trace-even}
\begin{equation}
\Tr(\widehat{U_\Phi^p}(z)) =-|a|^p\left(\widetilde z^p+\widetilde z^{-p}\right) + 2s_p(|a|^p-1).
\end{equation}
Since $\det \widehat U_p^p(z)=1$, we may write the eigenvalues as $e^{\pm i\omega}$. Hence
\begin{equation}\label{eq:even:Im:1}
\cos\omega =-\frac{|a|^p}{2}\left(\widetilde z^p+\widetilde z^{-p}\right) + s_p(|a|^p-1).
\end{equation}

Since $z\in\overline{\mathcal A(e^{\sigma/p})}$ and $|\widetilde z|=|z|$, we have $e^{-\sigma}\leq |\widetilde{z}^p|\leq e^\sigma$.
Write
\begin{equation}
\widetilde z^p=s_p e^{u+i\theta},
\qquad
|u|\leq \sigma.
\end{equation}
Then
\begin{equation} \label{anothertrigid}
\frac12(\widetilde z^p+\widetilde z^{-p})= s_p \cosh(u+i\theta) = s_p\left(\cosh u\cos\theta+i\sinh u\sin\theta\right).
\end{equation}
Substitute in \eqref{eq:even:Im:1} to obtain
\begin{equation}
\cos\omega =-s_p\left( 1-|a|^p + |a|^p (\cosh u \cos \theta + i \sinh u \sin \theta) \right).
\end{equation}

Multiplication by $-s_p\in\{+1,-1\}$ only changes $\omega$ by a possible shift by $\pi$, e.g.\ \eqref{addformula}, and hence does not change its imaginary part. Thus without loss of generality, we may assume that $s_p=-1$. 

Define
\begin{equation}\label{def:Y-even}
Y_e(u,p):=2\sinh^{-1}\left(|a|^{p/2}\sinh\left(u/2\right)\right),
\end{equation}
which is increasing, Observe that, using $\cosh u=1+2\sinh^2(u/2)$, we obtain
\begin{equation}\label{eq:even:coshY}
\cosh(Y_e(u,p))= 1+|a|^p(\cosh u-1).
\end{equation}

We now claim that
\begin{equation}\label{eq:even:elliple-0}
\frac{(\Re\cos\omega)^2}{\cosh^2( Y_e(u,p))}+\frac{(\Im\cos\omega)^2}{\sinh^2 (Y_e(u,p))}\leq 1,
\end{equation}
and postpone the proof to the end of our argument. 
Since $|u|\leq \sigma$, 
\begin{equation}
\cosh^2(Y_e(u,p))\leq \cosh^2(Y_e(\sigma,p)) \quad \text{ and }\quad \sinh^2(Y_e(u,p))\leq \sinh^2(Y_e(\sigma,p)).
\end{equation}
Hence, claim \eqref{eq:even:elliple-0} yields
\begin{equation}\label{eq:even:elliple-sigma}
\frac{(\Re\cos\omega)^2}{\cosh^2( Y_e(\sigma,p))}+\frac{(\Im\cos\omega)^2}{\sinh^2 (Y_e(\sigma,p))}\leq 1.
\end{equation}
Let  $\omega=x+iy$. We claim that \eqref{even-Im-bound} is valid. Suppose on the contrary, that
\begin{equation}
|y|>Y_e(\sigma,p) \quad \iff \quad \cosh^2 y>\cosh^2(Y_e(\sigma,p)) \text{ and }\sinh^2 y>\sinh^2(Y_e(\sigma,p))
\end{equation}
and in \eqref{eq:even:elliple-sigma}, use \eqref{addformula}
to get
\begin{equation}
\cos^2 x\frac{\cosh^2 y}{\cosh^2(Y_e(\sigma,p))}+\sin^2 x\frac{\sinh^2 y}{\sinh^2(Y_e(\sigma,p))}>\cos^2 x+\sin^2 x=1
\end{equation}
This contradicts \eqref{eq:even:elliple-sigma} and thus \eqref{eq:even:elliple-0}. Hence
\begin{equation}
|\Im\omega|= |y| \leq Y_e(\sigma,p)=2\sinh^{-1}\left(|a|^{p/2}\sinh(\sigma/2)\right),
\end{equation}
as desired.

It remains to prove \eqref{eq:even:elliple-0}. Define 
\begin{equation}
W(u)=\cosh u\cos\theta+i\sinh u\sin\theta
\end{equation}
and observe that $W(u)$ defines an ellipse (on the complex plane) as
\begin{equation}
\frac{(\Re W(u))^2}{\cosh^2 u}+\frac{(\Im W(u))^2}{\sinh^2 u}=\cos^2\theta+\sin^2\theta= 1.
\end{equation} 
This ellipse is centered on the origin, has focal points on the real axis $(1,0)$ and $(-1,0)$ because $\cosh^2 u-\sinh^2 u=1$. Thus $W(u)$ satisfies
\begin{equation} \label{eq:even:ellipse-W}
|W(u)-1|+|W(u)+1|=2\cosh u.
\end{equation}
Next, define
\begin{equation}
F(u)=1-|a|^p+|a|^pW(u)
\end{equation}
and observe that, by \eqref{anothertrigid}, $F(u)=\cos\omega$ when $s_p=-1$. Then proving \eqref{eq:even:elliple-0} is equivalent to proving
\begin{equation}\label{eq:even-ellipse-F}
|F(u)-1|+|F(u)+1|\leq 2\cosh(Y_e(u,p))
\end{equation}
Indeed, first
\begin{equation}\label{eq:even-ellipse-F-}
|F(u)-1|=|a|^p |W(u)-1|.
\end{equation}
and 
\begin{equation}\label{eq:even-ellipse-F+}
|F(u)+1|\leq 2(1-|a|^p)+|a|^p|W(u)+1|
\end{equation}
Thus, use \eqref{eq:even-ellipse-F-} and \eqref{eq:even-ellipse-F+} to obtain
\begin{align}
|F(u)-1|+|F(u)+1| &\leq  2(1-|a|^p)+|a|^p\left(|W(u)-1|+|W(u)+1|\right) \notag\\
& = 2(1-|a|^p)+2|a|^p\cosh u \notag\\
& = 2\cosh(Y_e(u,p)),
\end{align}
where \eqref{eq:even:ellipse-W} and \eqref{eq:even:coshY} were used in the last two steps.
This yields \eqref{eq:even-ellipse-F} and finishes the proof.
\end{proof}

\subsection{Deformation of a contour integral} \label{sec53}
Here we will mostly comment on the changes needed for even $p$ and otherwise refer back to Section~\ref{deformation} for the odd case.

Let $p$ be even, $0<|a|<1$ and $\sigma>0$ such that $\sigma < 2 \cosh^{-1}(|a|^{-p/2})$, or, equivalently $|a|^p \cosh^2 \left( \frac{\sigma}{2} \right) < 1$. 
In this case, the spectrum of $\widehat{U_\Phi^p}(z)$ is not simple on the whole closed annulus $\overline{\mathcal{A}(e^{\sigma/p})}\subset \overline{\mathcal{A}(\rho_{e}(|a|))}$, (where $\rho_{e}$ is defined in \eqref{def:Z}) because of the degeneracy at the points in $\mathcal Z_p(a)$ on the unit circle. However, the two contours
\begin{equation}
|z|=e^{\sigma/p}
\qquad\text{and}\qquad
|z|=e^{-\sigma/p}
\end{equation}
stay away from $\mathcal Z_p(a)$. By Lemma~\ref{lem:even-gap-away}, the gap between the two eigenvalues is uniform on these two contours.

This follows from observing that 
\begin{equation}
|\tilde{z}^p-s_p|\geq \left||\tilde{z}|^p-1\right|\geq 1-e^{-\sigma}
\end{equation}
on the contours $|z|=e^{\pm\sigma/p}$. Thus Lemma~\ref{lem:even-gap-away} applies with $\delta=1-e^{-\sigma}$.

For each (fixed) $z$ on the deformed contours, denote the two eigenvalues of
$\widehat U_\Phi^p(z)$ by (these eigenvalues are defined pointwise on the contour and they are written in this format because of $\det\widehat{U_\Phi^p}(z)=1$)
\begin{equation}
\lambda_\pm(z)=e^{\pm i\omega(z,a)}.
\end{equation}

We can follow \eqref{specrep} to \eqref{diagonalization} as before, and, in particular conclude \eqref{Ppmbound}.

By Lemma \ref{lem:even-Im-bound}, for all $z\in\overline{\mathcal A(e^{\sigma/p})}$,
\begin{equation}
|\Im\omega(z,a)|\leq Y_e(\sigma,p),
\end{equation}
where $Y_e(\sigma,p)$ is defined in \eqref{def:Y-even}.
By \eqref{diagonalization}, on the two deformed contours,
\begin{equation}
\left\|\left(\widehat{U_\Phi^p}(z)\right)^n\right\|\leq C(\sigma,p)e^{n Y_e(\sigma,p)}.
\end{equation}

We can now derive \eqref{normUpnbound} as before.

Using $|z|=e^{\sigma/p}$ on the deformed contour and first assuming $j\geq k$, we get
\begin{equation}
\|U_\Phi^{pn}(j,k)\|\leq C(\sigma,p)e^{n Y_e(\sigma,p)}e^{-(j-k)\sigma/p}.
\end{equation}

Similarly, if $j <k$, we deform the contour to $|z|=e^{-\sigma/p}$ and obtain
\begin{equation}
\|U_\Phi^{pn}(j,k)\|\leq C(\sigma,p)e^{n Y_e(\sigma,p)}e^{-(k-j)\sigma/p}.
\end{equation}

Therefore, for all $j, k, n \in \Z$,
\begin{equation} \label{pevenbound}
\|U_\Phi^{pn}(j,k)\|\leq C(\sigma,p)\exp\left[-\frac{\sigma}{p}\left(|j-k|-\frac{p}{\sigma}Y_e(\sigma,p)|n|\right)\right],
\end{equation}
where the extension to negative $n$ was done as before. We also use Lemma~\ref{scaling} as before. 
This finishes the proof of Theorem~\ref{thm:LR-bound} for even $p$.

\begin{appendix}
\section{The probability of being outside the LR-light cone} \label{sec:lightcone}
In this appendix we prove Corollary \ref{coro:tail}.

For the model $U_\Phi$ defined in \eqref{def:U-Phi} and \eqref{def:Phi}, assume that $0<|a|<1$. We use $\kappa(p)$ as defined in \eqref{def:kappa}.

Recall that by Theorem \ref{thm:LR-bound}, for every
\begin{equation}\label{eq:tail:sigma}
0<\sigma<\kappa(p)\cosh^{-1}(|a|^{-\frac{p}{\kappa(p)}}),
\end{equation}
there exists a constant $C(\sigma,p)<\infty$ such that
\begin{equation}\label{Tail:LR}
\|U_\Phi^n(j,k)\|\leq C(\sigma,p)e^{-\frac{\sigma}{p}(|j-k|-v(\sigma,p)|n|)},
\end{equation}
for all $j,k,n\in\Z$, where
\begin{equation}\label{eq:v-tail}
v(\sigma,p):=\frac{\kappa(p)}{\sigma}\sinh^{-1}\left(|a|^{\frac{p}{\kappa(p)}}\sinh\left(\frac{\sigma}{\kappa(p)}\right)\right).
\end{equation}
In particular, $v_{\LR}^{\inf}(U_\Phi)=|a|^{\frac{p}{\kappa(p)}}$.

Given $\epsilon>0$, let
\begin{equation}
v_\epsilon:=v_{\LR}^{\inf}(U_\Phi)+\epsilon=|a|^{\frac{p}{\kappa(p)}}+\epsilon.
\end{equation}
For an initial state $|j,+\rangle$ and an integer $n$, we are interested in bounding
\begin{equation}\label{def:P-tail}
P_{j,n,\epsilon}:=\sum_{k:\ |j-k|>v_\epsilon|n|}p_{j,n}(k), \text{ where} \ p_{j,n}(k):=\sum_{s\in\{+,-\}}|\langle k,s| U_\Phi^n|j,+\rangle|^2, \quad k\in\Z.
\end{equation}
$P_{j,n,\epsilon}$ in \eqref{def:P-tail} is the probability that a particle initially localized in the state $|j,+\rangle$ lies outside the Lieb--Robinson light cone of radius $v_\epsilon|n|$. The freedom in the parameter $\sigma$ can be used to optimize the exponential decay rate furnished by \eqref{Tail:LR}. This gives the sharp leading coefficient of $\epsilon^{3/2}$.

It follows from \eqref{Tail:LR} that
\begin{equation}\label{eq:p-pointwise}
p_{j,n}(k)\leq \|U_\Phi^n(k,j)\|^2
\leq C(\sigma,p)^2e^{-\frac{2\sigma}{p}(|j-k|-v(\sigma,p)|n|)}.
\end{equation}
Consequently,
\begin{equation}
P_{j,n,\epsilon}
\leq C(\sigma,p)^2e^{-\frac{2\sigma}{p}(V_\epsilon-v(\sigma,p))|n|}
\sum_{k:\ |j-k|>V_\epsilon|n|}
e^{-\frac{2\sigma}{p}(|j-k|-V_\epsilon|n|)}
\leq \tilde{C}_\epsilon  e^{-\Gamma_\epsilon(\sigma)|n|},
\label{main1}
\end{equation}
where
\begin{equation}\label{def:Gamma-epsilon}
\tilde{C}_\epsilon =\frac{2C(\sigma,p)^2}{1-e^{-2\sigma/p}} \ \text{ and }\ \Gamma_\epsilon(\sigma):=\frac{2\sigma}{p}\bigl(V_\epsilon-v(\sigma,p)\bigr).
\end{equation}
Thus, it remains to maximize $\Gamma_\epsilon(\sigma)$ over the admissible values of $\sigma$ in \eqref{eq:tail:sigma}.
Set $x:=\frac{\sigma}{\kappa(p)}>0$ and use \eqref{eq:v-tail} to observe  that
\begin{equation}
\Gamma_\epsilon(\sigma(x))=\frac{2\kappa(p)}{p}\left(V_\epsilon x-\sinh^{-1}\left(|a|^{\frac{p}{\kappa(p)}}\sinh x\right)\right)
\end{equation}

The Taylor expansions
\begin{equation}
\sinh x=x+\frac{x^3}{6}+\mathcal{O}(x^5),
\qquad
\sinh^{-1}x=x-\frac{x^3}{6}+\mathcal{O}(x^5)
\end{equation}
give
\begin{equation}\label{eq:h-expansion}
\sinh^{-1}\left(|a|^{\frac{p}{\kappa(p)}}\sinh x\right)=|a|^{\frac{p}{\kappa(p)}}x+\frac{A_p}{3}x^3+\mathcal{O}(x^5) \text{ as }x\to 0^+.
\end{equation}
where
\begin{equation}
A_p:=\frac{1}{2}\left(|a|^{\frac{p}{\kappa(p)}}-|a|^{\frac{3p}{\kappa(p)}}\right)
\end{equation}
Thus, 
\begin{equation}\label{def:gamma-opt} 
\Gamma_\epsilon(\sigma(x))=\frac{2\kappa(p)}{p}\left(\epsilon x-\frac{A_p}{3}x^3+\mathcal{O}(x^5)\right).
\end{equation}
Observe that the leading term is maximized when $\epsilon-A_p x^2=0$ where the unique maximizer is
\begin{equation}\label{eq:tail:x-epsilon}
x_\epsilon=\sqrt{\frac{\epsilon}{A_p}}.
\end{equation}
Note that for sufficiently small $\epsilon>0$, $\sigma=\kappa(p)x_\epsilon$ can be chosen so that \eqref{eq:tail:sigma} is satisfied.
 
Finally, substituting \eqref{eq:tail:x-epsilon} in \eqref{def:gamma-opt} yields
\begin{align}
\gamma_\epsilon
&=\frac{2\kappa(p)}{p}\left(\epsilon x_\epsilon-\frac{A_p}{3}x_\epsilon^3+\mathcal{O}(x_\epsilon^5)\right) \notag\\
&=\frac{4\kappa(p)}{3p\sqrt{A_p}}\epsilon^{3/2}+\mathcal{O}(\epsilon^{5/2}),
\end{align}
which proves \eqref{eq:gamma-asymptotics} and finishes the proof of Corollary \ref{coro:tail}.

\section{Asymptotic versus infimal LR-velocities} \label{sec:vAsyvLR}
In this appendix, we clarify the relation between the asymptotic velocity and the infimal LR-velocity of a banded quantum walk. We first show that the asymptotic velocity is always bounded above by the infimal LR-velocity (Lemma \ref{lem:vAsyvLR}). We then construct an inhomogeneous shift--coin walk for which this inequality is strict (Section \ref{sec:counterexample}).

\subsection{LR-velocities are upper bounds for the asymptotic velocity}
\begin{lemma}\label{lem:vAsyvLR}
Let $W$ be a banded quantum walk on
$\ell^2(\mathbb Z;\mathbb C^2)$, then
\begin{equation}\label{eq:vAsyvLR}
v_{\Asy}(W)\leq v_{\LR}^{\inf}(W).
\end{equation}
\end{lemma}

\begin{proof}
For every state $\psi\in D(Q)$ and $n\in\Z$, since $W^n$ is unitary, we have
\begin{equation}
\|Q W^n\psi\|\leq \|[Q,W^n]\psi\|+\|Q\psi\|.
\end{equation}
Then, because $\psi\in D(Q)$, we obtain
\begin{equation}\label{eq:vAsy-bound}
v_{\Asy}(W)=\sup_{\substack{\psi\in D(Q)\\ \|\psi\|=1}}\limsup_{n\to\infty}\frac{1}{n}\|QW^n\psi\| \leq \limsup_{n\to\infty}\frac{1}{n}\|[Q,W^n]\|.
\end{equation}
Let $v$ be a LR-velocity of $W$, that is, there exist $C<\infty$ and $c>0$ such that
\begin{equation}\label{vAsy-vLR:LR}
\|W^n(j,k)\|\leq C e^{-c(|j-k|-v|n|)}
\end{equation}
for all $j,k,n\in \Z$. 

For $k\in\Z$, let $(W^n)_k$ be the $k$-th diagonal of $W^n$, i.e, (on $\ell^2(\Z)\otimes \C^2$ )
\begin{equation}
(W^n)_k:=\sum_{\ell\in\Z}|\ell+k\rangle\langle \ell| \otimes W^n(\ell+k,\ell).
\end{equation}
Direct inspection shows that
\begin{equation}\label{vAsy-vLR:norm-W-k}
\|(W^n)_k\|_\mathcal{H}=\sup_{\ell\in\Z}\|W^n(\ell+k,\ell)\|_{\C^2}\leq^{(\ref{vAsy-vLR:LR})} Ce^{-c(|k|-v|n|)}
\end{equation}
and 
\begin{equation}\label{vAsy-vLR:Q-Wn}
\left[Q, (W^n)_k\right]=k(W^n)_k.
\end{equation}

Now, for any given $\epsilon>0$ and $n\in\Z$, let 
\begin{equation}\label{def:Kn}
K_{n,\epsilon}:=\frac{v|n|}{1-\epsilon}.
\end{equation}
Then, write $W^n$ as a sum of two matrices as follows
\begin{equation}\label{Wn:An+Bn}
W^n=A_{n,\epsilon}+B_{n,\epsilon}
\end{equation}
where $A_{n,\epsilon}$ is the banded matrix part of $W^n$ with band radius at most $\lfloor K_{n,\epsilon}\rfloor$ as follows.
\begin{equation}
A_{n,\epsilon}=\sum_{|k|<K_{n,\epsilon}}(W^n)_k, \quad \text{ and }\quad B_{n,\epsilon}:=\sum_{|k|\geq K_{n,\epsilon}}(W^n)_k
\end{equation}
Next, we bound $\|[Q,A_{n,\epsilon}]\|$ and  $\|[Q,B_{n,\epsilon}]\|$. Using \eqref{vAsy-vLR:Q-Wn}, we obtain
\begin{equation}\label{vAsy-vLR-Q-An1}
\|[Q,A_{n,\epsilon}]\|\leq K_{n,\epsilon}\|A_{n,\epsilon}\| \leq K_{n,\epsilon}+K_{n,\epsilon}\|B_{n,\epsilon}\|
\end{equation}
where we used $A_{n,\epsilon}=W^n-B_{n,\epsilon}$ and that $W^n$ is unitary. Then 
\begin{eqnarray}
K_{n,\epsilon}\|B_{n,\epsilon}\|&\leq^{(\ref{vAsy-vLR:norm-W-k})}& C K_{n,\epsilon}\sum_{|k|\geq K_{n,\epsilon}}e^{-c(|k|-v|n|)} \notag\\
&\leq& C \lceil K_{n,\epsilon}\rceil \sum_{|k|\geq \lceil K_{n,\epsilon}\rceil }e^{-c\epsilon |k|} \notag\\
&=& 2C \lceil K_{n,\epsilon}\rceil\ e^{-c\epsilon \lceil K_{n,\epsilon}\rceil}\sum_{k=0}^\infty e^{-c\epsilon k} \notag\\
&\leq& 2C \left(\sup_{x>0}xe^{-c\epsilon x}\right)\frac{1}{1-e^{-c\epsilon}}=:C_{1,\epsilon} \label{vAsy-vLR-KB}
\end{eqnarray}
Here, in the first-to-second step we used the fact that when $|k|\geq K_{n,\epsilon}=\frac{v|n|}{1-\epsilon}$, we have $v|n|\leq (1-\epsilon)|k|$. Use \eqref{vAsy-vLR-KB} in \eqref{vAsy-vLR-Q-An1} to obtain
\begin{equation}\label{vAsy-vLR-Q-An} 
\|[Q,A_{n,\epsilon}]\|\leq \frac{v|n|}{1-\epsilon}+C_{1,\epsilon}
\end{equation}
Similarly,
\begin{equation}\label{vAsy-vLR-Q-Bn}
\|[Q,B_{n,\epsilon}]\| \leq C \sum_{|k|\geq K_{n,\epsilon}} k\ e^{-c(|k|-v|n|)}\leq 2C\sum_{k=0}^\infty k e^{-c\epsilon  k }=:C_{2,\epsilon}
\end{equation}
Thus, use \eqref{vAsy-vLR-Q-An} and \eqref{vAsy-vLR-Q-Bn} in \eqref{Wn:An+Bn} to obtain,
\begin{equation}
\|[Q,W^n]\|\leq \|[Q,A_{n,\epsilon}]\|+\|[Q,B_{n,\epsilon}]\| 
\leq \frac{v|n|}{1-\epsilon}+C_{1,\epsilon}+C_{2,\epsilon}
\end{equation}
Divide by $n$, take the limit  $n\to\infty$ then $\epsilon\to 0^+$ to see, by \eqref{eq:vAsy-bound} that $v_{\Asy}(W)\leq v$. Since $v$ was an arbitrary LR-velocity for $W$, taking the infimum over all LR-velocities proves \eqref{eq:vAsyvLR}.

\end{proof}
\subsection{An example with strict inequality}\label{sec:counterexample}
Here we provide an example of a quantum walk $U$ on $\ell^2(\Z;\R^2)$ for which $v_{\Asy}(U)$ is strictly less than $v_{\LR}^{\inf}(U)$. This example is motivated by the discussion in \cite[Section 2.1]{ARJS26}.

Let a symmetric bi-infinite sequence $(j_m)_{m\in\Z}\subset \Z$ be given such that
\begin{equation}\label{def:jm}
j_0=0<j_1<j_2\ldots,\ j_{-k}=j_k \text{ for all }k,\text{  and }|j_{m+1}-j_m| \to \infty \text{ as }|m|\to\infty.
\end{equation}
Consider the inhomogeneous shift-coin quantum walk $U=SC$ on $\ell^2(\Z)\otimes\C^2$, where $S$ is the conditional shift operator \eqref{def:S}, and $C$ is the coin operator acting locally on position $n\in\Z$ by the unitary matrix $C(n)$, where
\begin{equation}
C(n)=\idty_2 \text{ if }n\notin\{j_m,\ m\in\Z\}, \text{ and }
C(j_m)=\begin{bmatrix}
0 & 1\\ -1& 0
\end{bmatrix} \text{ for all }m\in\Z.
\end{equation}
That is, $C(n)$ is a perfect reflector on sites $(j_m)_{m\in\Z}$, and it is a perfect transmitter elsewhere.

It is easy to see that $v_{\Asy}(U)=0$ because $U$ decomposes as a direct sum of finite-size blocks, see the discussion in \cite[Section 2.1]{ARJS26}. We claim here that $v^{\inf}_{\LR}(U)=1$. Hence, we prove that
\begin{equation} \label{AsylessLR}
0=v_{\Asy}(U)<v^{\inf}_{\LR}(U)=1.
\end{equation}

First observe that for any given $m,j,k\in\Z$ we have
\begin{equation}
\|U^{m}(j,k)\|=0 \text{ for }|j-k|>|m| \quad\text{and}\quad \|U^{ m}(j,k)\|\leq 1 \text{ for }|j-k|\leq |m|,
\end{equation}
see the corresponding argument in Section~\ref{sec:LRvel}.
Thus
\begin{equation}
\|U^m(j,k)\|\leq e^{-(|j-k|-|m|)}
\end{equation}
for all $j,k,m\in\Z$. Hence, $v=1$ is a LR-velocity, i.e., 
\begin{equation}
v_{\LR}^{\inf}(U)\leq 1.
\end{equation}
In the following, we will show the other direction: $v_{\LR}^{\inf}(U)\geq 1$. This means that $v_{\LR}^{\inf}(U)= 1$.

Assume on the contrary that $v_{\LR}^{\inf}(U)<1$. This implies that there exist constants $C<\infty$, $c>0$, and $v\in[0,1)$ such that
\begin{equation}\label{LR-assum}
\|U^n(j,k)\|\leq C e^{-c(|j-k|-v|n|)}
\end{equation}
for all $j,k,n\in\Z$.

For any given $m\in\Z_+$, choose a cavity $(j_{\ell_m}, j_{\ell_m+1})$ such that 
\begin{equation}
|j_{\ell_m}- j_{\ell_m+1}|\gg m. 
\end{equation}
In fact, we can choose the length of the cavity to be greater than $2m$. 

Note that there are infinitely many such cavities because their sizes are divergent to infinity, see \eqref{def:jm}. Then for $j:= j_{\ell_m}+1$, the sites $j$ and $j+m$ are both in the same cavity, and, due to the perfect transmitters,
\begin{equation}\label{j-j+m}
U^m|j,+\rangle=|j+m,+\rangle.
\end{equation} 
This implies that $\|U^m(j,j+m)\|=1$ (``$\le$'' follows from \eqref{Normfact} and ``$\geq$'' from \eqref{j-j+m}).

Thus, by the assumption \eqref{LR-assum}, we have
\begin{equation}\label{!!}
\|U^m(j+m,j)\|=1\leq C e^{-c(|m|-v|m|)}=C e^{-c(1-v)|m|}, \ \ v\in[0,1).
\end{equation}
Since $m$ is arbitrary, it can be chosen so that the RHS of \eqref{!!} is smaller than one, which is a contradiction. So there is no LR-velocity that is smaller than 1. 

This completes the proof of \eqref{AsylessLR}.

\end{appendix}


\bibliographystyle{abbrvArXiv}
\bibliography{References.bib}
\end{document}